\documentclass[aps,prb,twocolumn,showpacs,amsmath,amssymb,superscriptaddress,floatfix]{revtex4-1}
\usepackage{graphicx}
\usepackage{dcolumn} 
\usepackage{bm} 
\usepackage{amsfonts}
\usepackage{amsmath}
\usepackage{ulem}
\usepackage{color}
\usepackage{hyperref}
\usepackage{ulem}

\usepackage{subfiles}

\newcommand{\sign}{sgn}
\newcommand{\Vector}[1]{\ensuremath{\left( \begin{array}{c} #1 \end{array} \right)}}
 
\begin{document}

\title{Electron transport in  quantum channels with  spin-orbit interaction: \\ Effects of the sign of the Rashba coupling and  applications to  nanowires}

\author{Leonid Gogin}
\affiliation{Dipartimento di Scienza Applicata e Tecnologia del Politecnico di Torino, I-10129 Torino, Italy}


\author{Fausto Rossi}
\affiliation{Dipartimento di Scienza Applicata e Tecnologia del Politecnico di Torino, I-10129 Torino, Italy}

\author{Fabrizio Dolcini}
\email{fabrizio.dolcini@polito.it}
\affiliation{Dipartimento di Scienza Applicata e Tecnologia del Politecnico di Torino, I-10129 Torino, Italy}

\begin{abstract}
We investigate the effects of the sign of the Rashba spin-orbit coupling (RSOC) on electron transmission through a single-channel Nanowire (NW) in the quantum coherent regime. We show that, while  for a finite length NW with homogeneous RSOC contacted to two electrodes   the sign of its RSOC does not affect electron transport, the situation can be quite different in the presence of an inhomogeneous  RSOC and a  magnetic field   applied along the NW axis.  By analyzing transport across an interface between  two regions of different RSOC we find that, if the two regions have equal RSOC signs, the transmission within the magnetic gap energy range  is almost perfect,  regardless of the ratio of the spin-orbit energies to the Zeeman energy. In contrast, when the two regions have opposite  RSOC signs and are Rashba-dominated, the transmission gets suppressed. Furthermore, we discuss the implementation on a realistic NW setup where two RSOC regions are realized with suitably coupled gates separated by a finite distance. We find that the low-temperature NW conductance    exhibits a crossover from a short distance behavior that strongly depends on the relative RSOC sign of the two  regions to a large distance oscillatory behavior   that is  independent of such relative sign.  We are thus able to identify the conditions where  the NW conductance  mainly depends  on the sign of the RSOC and the ones where  only the RSOC magnitude  matters.

\end{abstract}

%

%
\maketitle
%
%


\section{Introduction}
Over the last 10 years most of research on semiconductor Nanowires (NWs) with  Rashba spin-orbit coupling (RSOC) has   focussed on the search for Majorana quasi-particles, which are believed to provide the building block for topologically protected quantum information\cite{dassarma_2010,vonoppen_2010,alicea_review, aguado_review,prada-review,valkov-review}. Although signatures compatible with these exotic quasi-particles have been found in NWs proximized by superconducting films\cite{kouwenhoven_2012,liu_2012,heiblum_2012,xu_2012,defranceschi_2014,marcus_2016,marcus_science_2016,kouwenhoven_2018}, the race to the evidence of Majorana states has somewhat  overshadowed 
 other interesting potentialities of NWs in quantum technologies.  

Indeed NWs   reaching   lengths of various $\mu {\rm m}$ and exhibiting quantum coherent transport are nowadays  fabricated both in the clean ballistic\cite{kouwenhoven_2018,kallaher_2010,kouwenhoven_nanolett_2013,kouwenhoven_nanolett_2016,shaepers_nanolett_2016,xu_2016,kouwenhoven_2017a,kouwenhoven_2017b,DeFranceschi_2018,kouwenhoven_2019a,kouwenhoven_2019b} and in the diffusive regime\cite{schaepers_2020}, and are also  realized in suspended geometries\cite{heiblum_2012,giazotto_2019},   in arrays\cite{kouwenhoven_2019b,bakkers_2012,philipose_2019} and in networks\cite{kouwenhoven_2019a}.     Moreover,  they can be used as flexible substrates for
hybrid epitaxial growth on selected facets in order to design  heterostructures with ferromagnets and superconductors\cite{krogstrup_2020}.  Also, NWs represent an extremely versatile and tunable platform for nanoelectronics since their conduction properties can be controlled both magnetically, e.g. by applying a magnetic field along the NW axis and thereby opening up  a gap in the spectrum, or electrically by controlling the RSOC through gate voltages. In particular, the   tremendous improvement in gating techniques   enables one  to  achieve a high control of  the RSOC\cite{sasaki_2013,micolich,sasaki_2017,das_2019,guo_2021,sasaki_2021}.   

Most of these efforts  have been devoted to enhance the {\it magnitude} of the RSOC  over a wide range of values\cite{gao_2012,slomski_NJP_2013,wimmer_2015,bercioux_review,nygaard_2016,sherman_2016,tokatly_PRB_2017,loss_2018,goldoni_2018,gao-review,lau_2021}. However,
when an electron travels through a given spin-orbit region, its spin polarization depends not only on the magnitude, but also on the  {\it sign} of the local spin-orbit coupling. Are there   any observable effects  that can be attributed to the sign? So far, this question has been addressed only in a few cases. Concerning the  Dresselhaus spin-orbit coupling, its sign reversal  has been investigated  in InGaAs rings\cite{nitta-frustaglia}, whereas in the case of the RSOC the problem has mainly been  addressed in two-dimensional quantum wells interfaced with other suitable materials\cite{kaindl_2005, slomski_2013,tsai_2018} or where bulk subbands are characterized by opposite RSOC signs\cite{wang-fu_2016}.
As far as one-dimensional single-channel NWs are concerned,  it has been shown that a spatially modulated RSOC with alternating sign can drive the transition from a metallic to an insulating state\cite{johannesson-japaridze_2011}. Moreover, when a NW is  exposed to a magnetic field along its axis and the RSOC magnitude is large,  the propagating states  in the magnetic gap exhibit a locking between propagation direction and spin, whose helicity is determined by the RSOC sign\cite{streda,depicciotto_2010,loss_PRB_2011,lutchyn_2012,loss_PRB_2017}. Nevertheless,  when the NW with homogeneous  RSOC  is contacted to two electrodes,  the electron transport turns out to depend only on the magnitude of the RSOC, and not on its sign, as we shall see below.  
Recent studies suggest that a different scenario may emerge  in the presence of inhomogeneous RSOC profile. A junction between two regions with opposite RSOC signs, for instance,   exhibits interesting  spectral and equilibrium properties at the interface, such as localized bound states\cite{loss_EPJB_2015,rossi-dolcini-rossi_EPJ} and orthogonal spin polarization\cite{rossi-dolcini-rossi_2020}, which look surprisingly similar to the ones of the topological phase.    
In experimental implementations, where the RSOC can be locally controlled by  gates,  these inhomogeneous   configurations can be    realized by coupling various gates along the NW, providing an additional knob to control  the electrical current flowing through the NW setup.  

The question arises whether in such inhomogeneous configurations the sign of the RSOC can lead to any observable effect on out of equilibrium properties. 
In this paper we investigate  this problem by analyzing the electron transport through a single-channel NW with inhomogeneous RSOC  in the coherent quantum limit. The paper is organized as follows. In Sec.\ref{sec-2} we present the model and describe the adopted method. Then, in Sec.\ref{sec-3} we  discuss the two necessary ingredients to observe the effects of the sign of the RSOC, namely the presence of a uniform magnetic field and the existence of at least two regions with different RSOC. In Sec.\ref{sec-4} we first perform a preliminary analysis of the transmission in the presence of a single interface separating two regions with different RSOC, finding a quite different behavior of the electrical conductance in the cases of equal and opposite RSOC signs. Then, by analyzing a more realistic configuration, where two differently gated portions of a NW are separated by a finite distance, we show that such lengthscale can modify the results obtained in the ideal limit of one single interface case. This enables us to identify the conditions where the sign of the RSOC affects the NW transport properties, and  the situations where only the RSOC magnitude matters.
Finally, in Sec.\ref{sec-6} we discuss our results and draw our conclusions.

\section{Model and method}
\label{sec-2}
 We  consider a one-dimensional  electron conduction channel directed along the $x$ direction, characterized by a  spatially varying RSOC profile and exposed to an external Zeeman magnetic field directed along its axis. The   second-quantized Hamiltonian reads
\begin{align} 
    \hat{\mathcal{H}}  = \int \hat{\Psi}^\dagger(x) \left(  \frac{p_x^2}{2 m^*}  \sigma_0 - \frac{\sigma_z}{2\hbar} \{ \alpha(x), p_x \}    - \sigma_x h_\perp  \right) \hat{\Psi}(x)  \label{H-inhomo}
\end{align}
where $m^*$ denotes the effective electron mass. In order to ensure   Hermiticity,  a half of the anticommutator between the momentum operator $p_x$ and the spatially inhomogeneous RSOC  $\alpha(x)$ has been introduced, as customary\cite{sherman_PRL_2007,brataas_2007,sherman_PRB_2013,sherman_PRB_2018}. Here $\sigma_x,\sigma_y, \sigma_z$ denote spin Pauli matrices, while $\sigma_0$ is the $2 \times 2$ identity. Moreover, $z$ is the direction of the RSOC effective magnetic field, whereas~$x$ is the direction of the actual magnetic field, characterized by an  energy coupling $h_\perp={\rm e} g \hbar B/4 m_e$, where ${\rm e}$ denotes the electron charge, $g$ the $g$-factor and $m_e$ the bare electron mass.  

We are interested in describing electron transport along the inhomogeneous channel. With respect to spin degenerate problems, the charge current operator exhibits an additional term related to the RSOC\cite{rashba_2004,dolcini-rossi_2018}
\begin{eqnarray}
    \hat{J}^c(x,t) &=& -  \frac{i{\rm e} \hbar}{2m^*} \left( \hat{\Psi}^\dagger(x,t) \partial_x \hat{\Psi}(x,t)- \partial_x\hat{\Psi}^\dagger(x,t) \, \hat{\Psi}(x,t) \right)  \nonumber \\
    & & -\frac{{\rm e}\alpha(x)}{\hbar} \hat{\Psi}^\dagger\sigma_z   \hat{\Psi} \quad,\label{J-inhomogeneous}
\end{eqnarray}
as can be deduced from the very continuity equation $\partial_t \hat{n}+\partial_x \hat{J}^c=0$ following from the Heisenberg Equation dictated by the  Hamiltonian (\ref{H-inhomo}), where the charge density is $\hat{n}= {\rm e}\hat{\Psi}^\dagger \hat{\Psi}^{}$.

\subsection{Relation with a magnetic texture problem}
\label{sec-mapping}
Before discussing any specific RSOC profile, we note a general property of the  Hamiltonian (\ref{H-inhomo}). By performing the transformation\cite{sanchez_2006,sherman_PRL_2007}
\begin{equation}\label{gauge-transf}
\hat{\Psi}(x)=\displaystyle e^{i \phi_{SO}(x)  \frac{\sigma_3}{2}} \, \hat{\Psi}^\prime(x)
\end{equation}
where  
\begin{equation}\label{phiSO-def}
\phi_{SO}(x) = \frac{2m^*}{\hbar^2} \int_{x_0}^x \alpha(x^\prime)\, dx^\prime 
\end{equation}
and $x_0$ is an arbitrarily  fixed reference point, the Hamiltonian  (\ref{H-inhomo}) is rewritten in the new fields $\hat{\Psi}^\prime(x)$ as
\begin{eqnarray} 
\hat{\mathcal{H}}  &= &\int {\hat{\Psi}}^{\prime \dagger}(x)\, \Bigg( \frac{p^2_x}{2 m^*} +U_{SO}(x)     -\label{Hmagn-texture} \\
& & -h_\perp \left( \cos[\phi_{SO}(x)] \sigma_x+ \sin[\phi_{SO}(x)] \sigma_y \right) \Bigg) {\hat{\Psi}^\prime}(x) \,dx \nonumber
\end{eqnarray} 
where 
\begin{equation}\label{USO-def}
U_{SO}(x)= -  \frac{{m^*}  \alpha^2(x)}{2\hbar^2}\quad.
\end{equation}
Various aspects   are noteworthy. First, in the absence of Zeeman field ($h_\perp=0$), while for a homogeneous problem Eq.(\ref{USO-def}) is just a mere energy constant and the spin-orbit coupling gets completely gauged out, in an inhomogeneous problem the spatial profile of the effective  scalar   potential $U_{SO}(x)$  does affect electron transmission, although in a spin-independent way.\cite{note-bc}  For instance, in the case of a piecewise constant RSOC,  $U_{SO}(x)$ acquires the form of  a potential step, a quantum well or a barrier, leading to an energy dependent transmission.  The second aspect is that, when the Zeeman term $h_\perp$ is additionally present,   the last term of Eq.(\ref{Hmagn-texture}) acquires the form of a magnetic texture in the  spin $x$-$y$ plane characterized by the rotation angle $\phi_{SO}(x) $ in Eq.(\ref{phiSO-def}), which represents the integrated local spin-orbit wavevector (multiplied by 2). In this case the role of the spin degree of freedom becomes not trivial because of the interplay between actual magnetic field and spin-orbit effective magnetic field. In the following, we shall occasionally refer to the mapping to Eq.(\ref{Hmagn-texture}) in order to interpret some of our results.

\subsection{Piecewise constant profile}
\begin{figure} 
\centering
    \includegraphics[width=\linewidth]{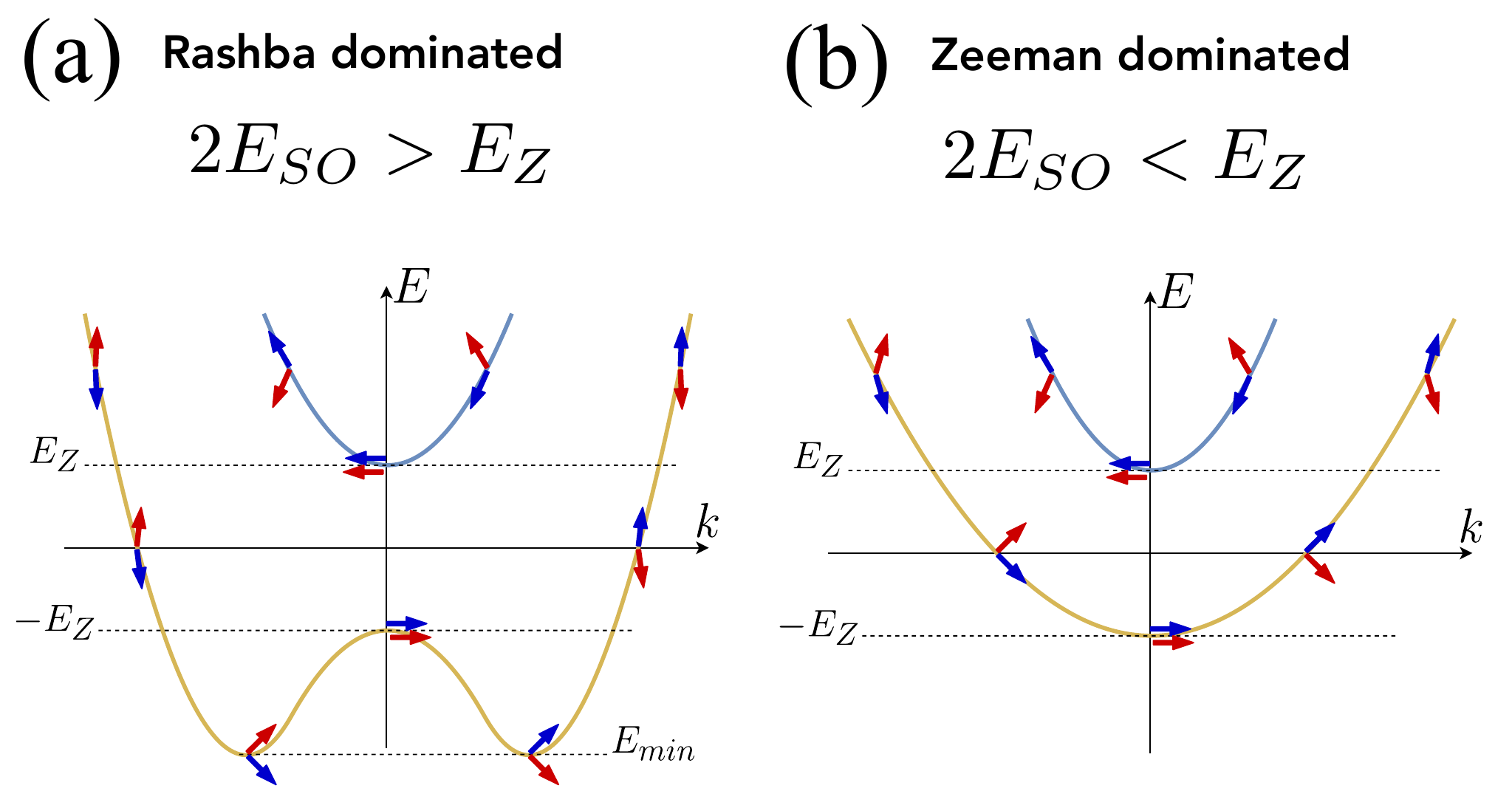}
    \caption{\small The local spectrum of the RSOC channel in each piecewise constant profile region.  (a) The case of the Rashba-dominated regime, where the two local  minima of the lower band are at $E_{min}=-E_{SO}(1+E_Z^2/4 E_{SO}^2)$; (b) The case of the   Zeeman-dominated regime.  In each panel the arrows indicate the direction of $k$-dependent spin orientation in the $x$-$z$ plane of the sheet, in the case of $\alpha>0$ (blue arrows) and  $\alpha<0$ (red arrows).  }
    \label{Fig1}
\end{figure}

Although for an arbitrary spatial profile the inhomogeneous problem  cannot be solved analytically, any profile $\alpha(x)$ can ultimately be approximated with a piecewise constant profile, where the solution can be build up  by suitably matching the electron field operator in each  region. It is thus worth recalling briefly the main physical ingredients characterizing each locally homogeneous  region. In a region with RSOC $\alpha_j$ one identifies two energy scales, namely the spin-orbit energy 
\begin{equation}\label{ESOj-def}
E_{SO,j}= \frac{m^*  \alpha_j^2}{2 \hbar^2}
\end{equation}
and the Zeeman energy 
\begin{equation}
E_{Z} = |h_\perp|
\end{equation}
which directly impact on the  local spectrum. Indeed it consists of two bands  (see Fig.\ref{Fig1}), separated by a gap $\Delta=2E_Z$ at $k=0$, where the lower band exhibits one minimum or two minima depending on whether the region is in  the Zeeman-dominated regime ($2E_{SO,j}<E_Z$) or in  the Rashba-dominated regime ($2E_{SO,j}>E_Z$).   
The classification of the eigenstates is in principle straightforward, although it   requires a little more care in practice. First, while the spectrum does not depend on the sign of the RSOC $\alpha_{j}$, the eigenstates do. Since this will play a role in the results to be presented below, in Fig.\ref{Fig1} we have highlighted  the $k$-dependent spin orientation in the $x$-$z$ plane of the local eigenstates, distinguishing the cases $\alpha>0$ and $\alpha<0$  by blue and red arrows, respectively. 
Furthermore, since in the quantum coherent limit the inhomogeneous solution is characterized by a given energy  $E$, in each region one has to identify all 4 wavevectors corresponding to such energy, retaining both   real wavevectors (propagating modes) and   complex wavevectors (evanescent modes). In particular, the expression of the evanescent modes, which turn out to play an important role when matching the wavefunctions in different regions,   leads to one further distinction between the weak Zeeman regime ($2E_{SO,j}<E_Z<4E_{SO,j}$) and the strong Zeeman regime ($E_Z>4E_{SO,j}$). After performing such  lengthy but straightforward  classification of all eigenstates, whose technical details and  results   are reported in Appendix \ref{AppA}, the general solution at energy $E$  is built up by matching the linear superpositions at each interface. Explicitly, at the $j$-th interface  located at position $x_{j}$ and separating two regions with RSOC values $\alpha_{j}$ (on the left) and $\alpha_{j+1}$ (on the right), the boundary conditions are
  \begin{equation}\label{bc-single-interface}
\left\{ \begin{array}{lcl}
        \hat{\Psi}(x_j^-) &=& \hat{\Psi}(x_j^+)\\ \\
          \partial_x \hat{\Psi}(x_j^-) &=& \partial_x \hat{\Psi}(x_j^+) -   \frac{i m^*}{\hbar^2} (\alpha_{j+1}-\alpha_{j}) \sigma_z  \hat{\Psi}(x_j)
    \end{array}\right. 
\end{equation}

Then, by applying the Scattering Matrix Formalism\cite{datta_book}, where the external regions act like the leads and the inhomogeneous profile determines the Scattering region, we compute the Scattering Matrix and determine the Transmission coefficient ${T}(E)$ for various inhomogeneous configurations. Details and examples are provided in Appendix \ref{AppB}. This enables us to determine how  the low temperature linear conductance 
\begin{align}\label{conductance}
    G &= \frac{{\rm e}^2}{h} {T}(E_F)
 \end{align}
depends on the  RSOC and the Zeeman field, allowing for an electrical and magnetic tuning of the transport properties.

\section{Two essential ingredients to observe effects of the RSOC sign}\label{sec-3}
Because our purpose is to analyze the effect  of the sign of the RSOC on transport properties, it is first worth pointing out some general conditions for this to be observed. 
We start by noting that  the transmission coefficient is completely independent of   a  {\it global} sign change of the RSOC profile, $\alpha(x)\rightarrow -\alpha(x)$. Indeed the Hamiltonian (\ref{H-inhomo}) with a profile $-\alpha(x)$ can be mapped back into the  one with a profile $+\alpha(x)$  through the transformation $\hat{\Psi}(x) \rightarrow  \sigma_x \hat{\Psi}(x)$. This  implies, for instance, that for a  finite length NW with a homogeneous RSOC  contacted two normal leads with vanishing RSOC, modelled with an  inhomogeneous profile  $\alpha(x)=\alpha \, \theta(d/2-|x|)$,  where~$\theta$ denotes the Heaviside function and $d$ is the NW length, the conductance is independent of $\mbox{sgn}(\alpha)$.   In particular, we note that the effect of Fano anti-resonances with vanishing  transmission occurring in such a case for suitably chosen parameters\cite{sanchez_2006,sanchez_2008,aguado_2015}  is completely insensitive to the sign of the  RSOC in the NW.   

Thus,  a necessary condition for the effects of the RSOC sign to be observed is that the sign of the RSOC in a portion of the profile $\alpha(x)$  
changes   with respect to the rest.
However,   it may not be sufficient. Compare for instance the uniform profile $\alpha(x)\equiv \alpha_0$ and the inhomogeneous profile $\alpha(x)=\alpha_0 \mbox{sgn}(x)$, where the sign on the left of the origin is changed with respect to the one on the right. Despite the abrupt sign change of the RSOC, in the absence of the Zeeman field  ($h_\perp=0$) the electron transmission is always perfect  and equal to the case of a uniform profile.
This can be  straightforwardly deduced from the mapping described in Sec.\ref{sec-mapping}, since the potential Eq.(\ref{USO-def}) reduces to a mere homogeneous constant $U_{SO}\equiv -   {m}^*  \alpha_0^2/2\hbar^2$. In contrast,  if $h_\perp \neq 0$, the   term in the second line of Eq.(\ref{Hmagn-texture}) is present and the phase $\phi_{SO}(x)$ in Eq.(\ref{phiSO-def}) depends on the sign of~$\alpha$. \\
 
From these remarks, we conclude that the two essential ingredients to observe effects of the RSOC sign on electron transport are i) a Zeeman field directed perpendicularly to the RSOC effective magnetic field and ii) the presence of at least {\it two} regions with different RSOC value. This is what we shall consider in the following, focussing on the intrinsic transmission properties of the inhomogeneous RSOC. The presence of spurious resistance due to non-ideal contacts with external electrodes can in principle be taken into account with the  method outlined in Appendix \ref{AppB}. However, it depends on the specific metals used as electrodes and on the adiabaticity of the contacts\cite{rainis-loss_PRB_2014}, and it goes beyond the purpose of this paper.

\section{The single interface problem}
\label{sec-4}
We start by analyzing  the transmission across an interface separating two regions with different RSOC, namely 
\begin{equation}\label{alpha-single}
\alpha(x)= \left\{ \begin{array}{lcl}  \alpha_{L}   & \mbox{for} & x <0 \\ & & \\ \alpha_{R}   & \mbox{for} & x >0 
\end{array} \right. \quad.
\end{equation}

\begin{figure} 
\centering
    \includegraphics[width=\linewidth]{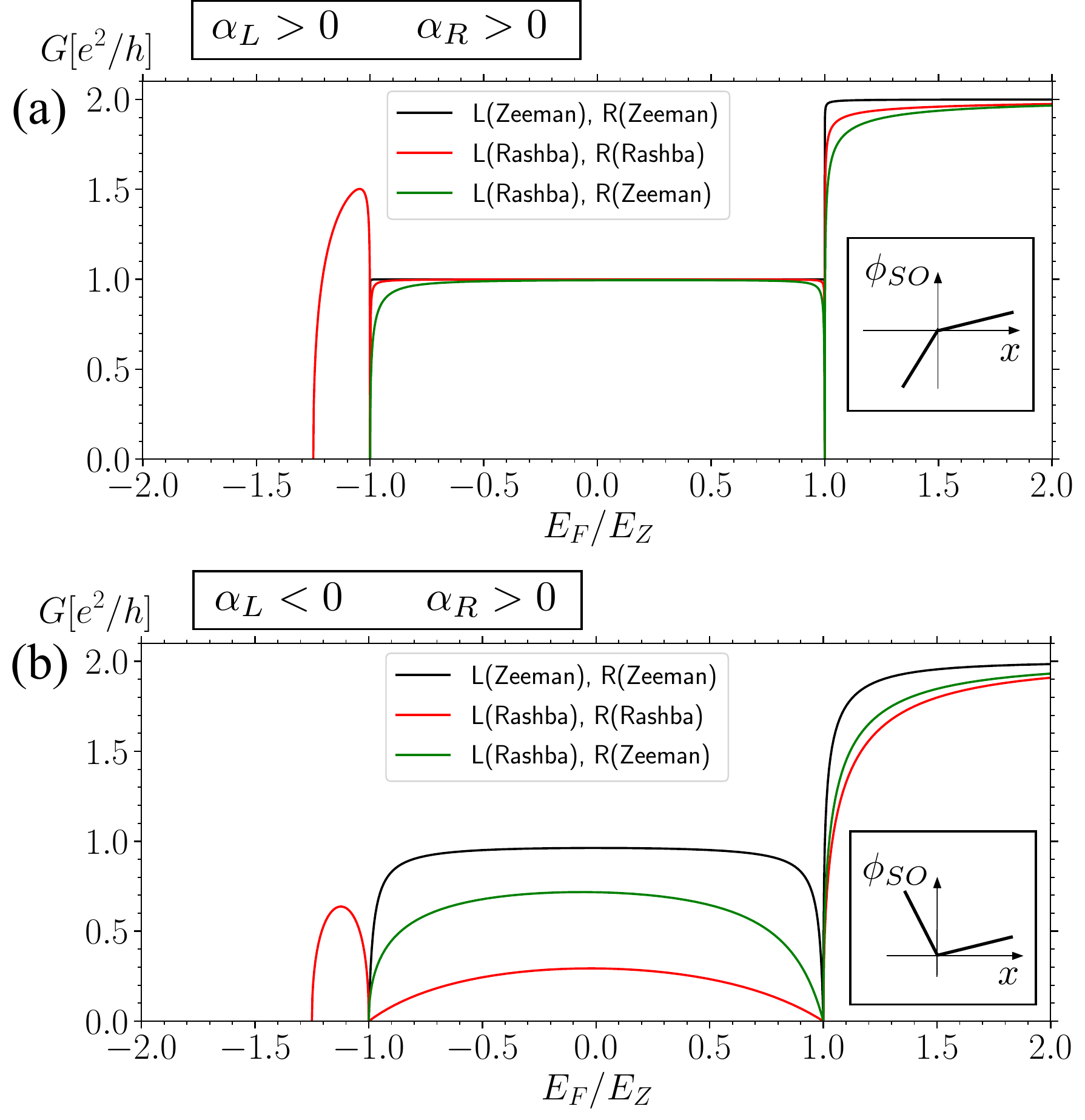}
    \caption{\small The conductance of the single interface problem [see Eq.(\ref{alpha-single})] is plotted in units of conductance quantum $G_0={\rm e}^2/h$ as a function of the Fermi energy $E_F$ (in units of the Zeeman energy $E_Z$) for various spin-orbit configurations.   In each panel the black curve refers to the case where both regions are Zeeman-dominated  ($2E_{SO,L}=0.3 E_Z$; $2E_{SO,R}=0.5 E_Z$), the red curve to the case where both regions are Rashba-dominated  ($2E_{SO,L}=2 E_Z$; $2E_{SO,R}=5 E_Z$), while  the green curve to the case where the left region is Rashba-dominated and the right region is Zeeman dominated ($2E_{SO,L}=3 E_Z$; $2E_{SO,R}=0.5 E_Z$). While panel (a) refers to the case where $\alpha_{L},\alpha_{R}>0$, panel (b) refers to the case of coupling changing sign across the junction, $\alpha_{L}<0$ and $\alpha_{R}>0$.  Insets: the spatial behavior of the spin-orbit angle (\ref{phiSO-def}) describing the rotation of the effective magnetic texture problem (\ref{Hmagn-texture}).}
    \label{Fig2}
\end{figure}

Depending on the strength of the RSOC, each interface side can be in the Rashba-dominated or in the Zeeman-dominated regime.  After solving the scattering problem, as  outlined in the Appendix~\ref{AppB}, we have determined the transmission coefficient and the conductance from Eq.(\ref{conductance}). 
Specifically, Fig.\ref{Fig2} shows the  conductance~$G$,  in units of the conductance quantum $G_0={\rm e}^2/h$, as a function of the Fermi energy~$E_F$, in units of the Zeeman energy~$E_Z$, for the three independent configurations: Zeeman-Zeeman, Rashba-Rashba, and Rashba-Zeeman\cite{nota-config}.  In particular, Fig.\ref{Fig2}(a) illustrates the case  where the RSOC has the same sign in both sides. Recalling that the energy range $|E_F|<E_Z$ corresponds to the magnetic gap (see Fig.\ref{Fig1}), we first note that, when both sides are in the Rashba-dominated regime (red curve), transmission is possible also in the energy range below the magnetic gap, where it can also exceed $G_0$. This is because  two propagating channels exist in the range   $E_{min}<E<-E_Z$, as one can see from the mexican-hat shaped lower band in Fig.\ref{Fig1}(a).  At the energies $E_F=\pm E_Z$ corresponding to the magnetic gap edges, the conductance of Fig.\ref{Fig2}(a) exhibits two sharp anti-resonance suppression cusps, an effect similar to the one found in Refs.[\onlinecite{sanchez_2008}] and [\onlinecite{aguado_2015}] in a NW coupled to two leads.  We shall comment about this aspect at the end of this section.  Apart from these features, one finds a practically perfect transmission, namely $G=G_0$ when the Fermi energy lies inside the magnetic gap, $|E_F|< E_Z$, (one propagating channel) and $G=2 G_0$ above the gap $E>E_Z$ (two propagating channels), regardless of the specific regime configuration.

A different scenario emerges when the RSOC takes opposite signs across the interface, as shown in  Fig.\ref{Fig2}(b).  In all configurations the conductance in the magnetic gap range $|E_F|<E_Z$ remains roughly symmetric with respect to the midgap energy value $E_F=0$. However, we note that, while the transmission  is practically perfect  when both sides are in the strong Zeeman regime (black curve),    it reduces  when one region enters the Rashba regime  (green curve) and even more when   both regions are Rashba-dominated   (red curve).  The different behavior of the conductance in the two panels (a) and (b) can be qualitatively understood through the mapping to the magnetic texture problem  Eq.(\ref{Hmagn-texture}). In the present case of the profile Eq.(\ref{alpha-single}) the rotation angle (\ref{phiSO-def}) of the magnetic field   acquires the form 
\begin{equation}
\phi_{SO}(x)= \frac{2m^*}{\hbar} \times \left\{ \begin{array}{lcl} \alpha_R x & \mbox{for} & x>0 \\ & & \\ \alpha_L x & \mbox{for} & x<0 \end{array} \right. 
\end{equation}
and its spatial behavior is depicted in the insets of the two panels of Fig.\ref{Fig2}. If $\alpha_L$ and $\alpha_R$ have the same sign [inset of panel (a)], the incoming wavefunction can easily adapt to the change of slope of $\phi_{SO}(x)$ from one side to the other by simply stretching or shrinking, since a coordinate rescaling $x\rightarrow x \,\alpha_R/\alpha_L$ would compensate for the slope change. This leads to a very high tranmission. In contrast, when $\alpha_L$ and $\alpha_R$ have opposite signs, an actual cusp appears in $\phi_{SO}$  [inset of panel (b)], which cannot be merely removed by an affine coordinate transformation. It is therefore more difficult for the incoming wavefunction spinor to re-adapt to the profile on the other side of the interface. As long as the values of the RSOC are small (both regions in the Zeeman dominated regime) this effect is negligible, but when both regions enter the Rashba-dominated regime the mismatch becomes important and the transmission is suppressed.

\begin{figure}[h!]
\centering
    \includegraphics[width=\linewidth]{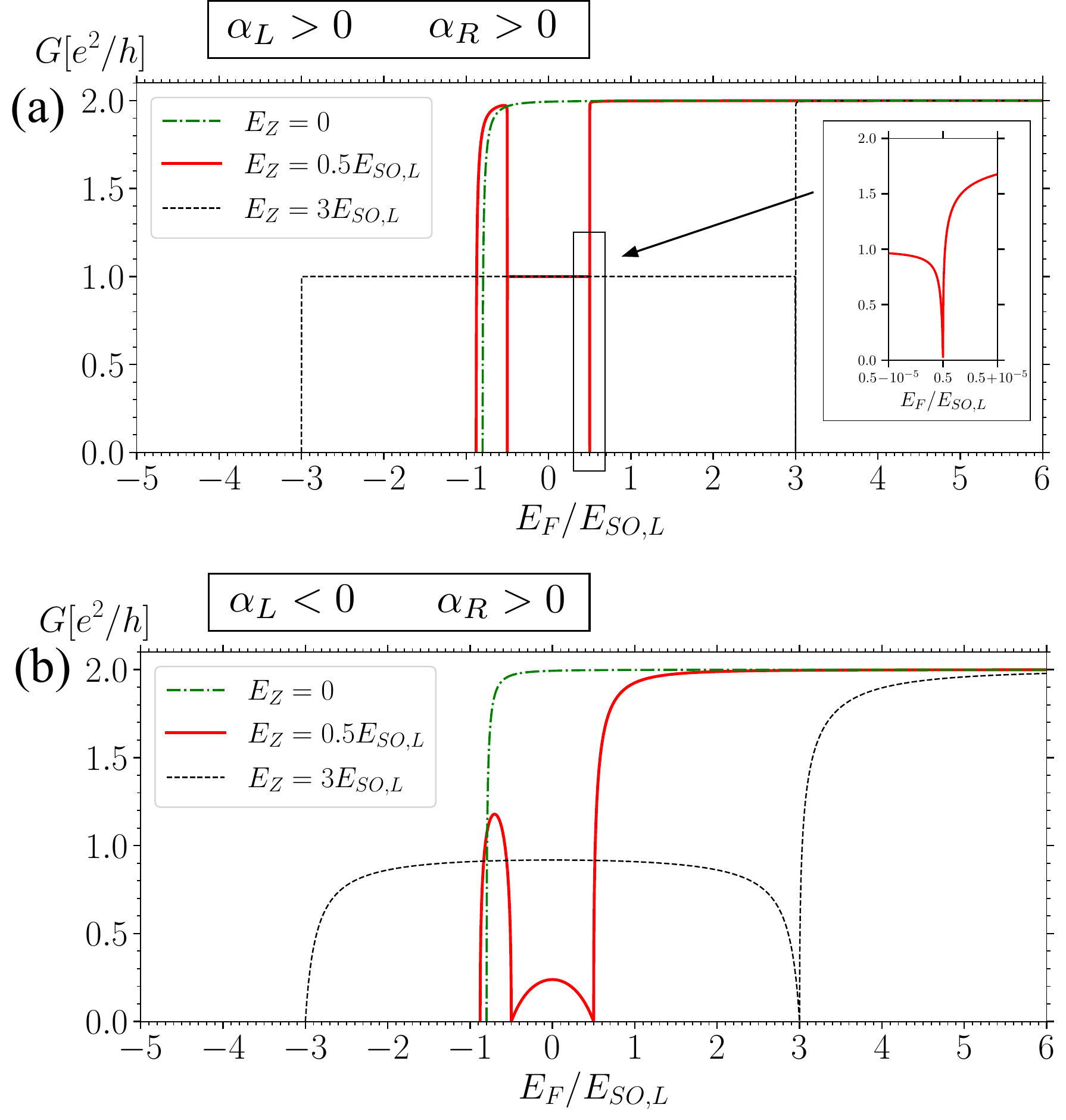}
    \caption{ \small   The single interface problem [see Eq.(\ref{alpha-single})]  with $E_{SO,R}=0.8 E_{SO,L}$. The conductance $G$ in units of the conductance quantum $G_0={\rm e}^2/h$  as function of the Fermi energy $E_F/E_{SO,L}$,  for different values of the Zeeman energy $E_Z$. (a) The case where the RSOC takes the same sign on both interface sides.  Note that, when both sides are in the Rashba-dominated regime (red curve) a peak of $2G_0$ conductance is present for $E_{min} < E< -E_Z$, i.e. for $-0.878 < E/E_{SO,L} < -0.5$, due to the presence of two propagating channels below the magnetic gap energy  (see Fig.\ref{Fig1}(a)). The inset on the right magnifies the sharp anti-resonance cusp occurring when the Fermi energy equals the upper boundary of the magnetic gap $E_F =+E_Z$. Similar behavior occurs at the lower boundary $E_F =-E_Z$.  In  this case the    Zeeman energy $E_Z$ only controls the number of propagating channels (0, 1 or 2) and the conductance varies discretely in practice. (b) The case where the RSOC has opposite signs across the interface. In this case the Zeeman energy also enables one to vary continuously the conductance over the entire energy range.   }
    \label{Fig3}
\end{figure}

The role of the RSOC sign can also be highlighted by considering the magnetic tuning of the conductance~$G$ in a Rashba interface where both sides are in the same regime.  As an illustrative example of such situation, Fig.\ref{Fig3}     describes  the Fermi energy dependence of $G$ for a junction with $E_{SO,R} / E_{SO,L} = 0.8$, for various values of the Zeeman energy $E_Z$. Specifically, panel(a) shows the case where the  RSOC takes the {\it same} sign across the interface ($\alpha_{L},\alpha_{R}>0$). When the magnetic field is absent  (green dash-dotted curve) the conductance vanishes  for $E_F<-E_{SO,R}$ and then rapidly increases to $2 G_0$ for $E_F>-E_{SO,R}$. This is  exactly the same energy dependence as   a spin-degenerate problem of transmission across a potential step $(-E_{SO,L}|-E_{SO,R})$, as   argued invoking the mapping described in Sec.\ref{sec-mapping}. In contrast, when a weak Zeeman field is introduced  (red solid curve), the conductance gets suppressed down to $G_0$ for $|E_F|<E_Z$ due to the magnetic gap opening that leaves only one propagating channel. Again,  very narrow cusps with vanishing conductance appear  at Fermi energies corresponding to the magnetic gap boundaries $E_F=\pm E_Z$ (one of them is highlighted in the inset on the left).   For a stronger Zeeman field (black dashed line)  both sides enter the Zeeman-dominated regime and the conductance acquires a step-like behavior characterizing the number of propagating channels (0, 1 or 2). Thus, in this case of equal RSOC signs, the Zeeman energy in practice controls only the number of such channels, i.e. the value at which the jump occurs.
Figure~\ref{Fig3}(b) instead describes the case of {\it opposite} RSOC sign across the interface, i.e. $\alpha_{L}<0$ and $\alpha_{R} >0$. For vanishing Zeeman field  the conductance is  insensitive to the sign of the RSOC, so that the green dash-dotted  curve is exactly equal to the one obtained in Fig.\ref{Fig3}(a). Again, exploiting the mapping (\ref{gauge-transf}) described in Sec.\ref{sec-mapping}, one can see that in such situation only the potential~(\ref{USO-def}) is present. In contrast, when the Zeeman field is introduced, the sign of the RSOC matters and the behavior strongly differs from panel~(a). The conductance varies continuously as a function of the Fermi energy. In this case the Zeeman energy $E_Z$ determines not only  the location of the suppression cusps at the magnetic gap edges, but also the magnitude of $G$ in the entire energy range. In particular,  when the NW is in the Rashba dominated regime (red curve), the conductance is significantly suppressed inside and below the magnetic gap. \\

{\it Anti-resonances in the conductance.} We conclude this section by a comment about the  anti-resonance cusps with vanishing conductance displayed in Figs.\ref{Fig2} and \ref{Fig3}. We emphasize that their  presence is not  due to the RSOC sign change across the interface, as they do exist both in the case of equal and opposite RSOC signs, and turn out to be much sharper in the former case [see inset of Fig.\ref{Fig3}(a)]. Although this effect is thus   generic  and not strictly related to the main focus of our paper, for the sake of completeness in the presentation of our results, a brief discussion about their origin is in order. We recall that the existence of vanishing conductance dips, typically close in energy to resonance peaks of perfect transmission,  is known to occur in the case of a NW    contacted to leads\cite{sanchez_2008,aguado_2015},  and is attributed to the fact that   the RSOC present in the centrally confined NW region leads to a spin misalignment 
of its bound states with respect to the outer leads with continuum spectrum, causing one bound state   to be strongly coupled to the continuum, while the other one is very weakly coupled.   
With respect to the configuration analyzed in Refs.[\onlinecite{sanchez_2008}] and [\onlinecite{aguado_2015}], our case of a single interface between two RSOC regions  exhibits two differences: i) here, within the energy gap,  there are no bound states at the interface\cite{nota-su-bs}; ii) the anti-resonances with vanishing transmission are always pinned at the boundaries $E=\pm E_Z$ of the magnetic gap. Their origin can be argued as follows. Within the energy range $|E|<E_Z$ of the magnetic gap, each side of the interface is characterized by propagating modes (see Fig.\ref{Fig1}), whose spin orientations lying in the $x$-$z$ spin plane are in general misaligned, because the RSOC  takes  different values  across the interface. However, one also has evanescent modes originating from the band spin-splitting in the magnetic gap  [see Fig.\ref{Fig8}(b) and Fig.\ref{Fig9}(b) and (c) in App.\ref{AppA}], whose spin orientation lies in the $x$-$y$ spin plane. 
On each interface side the wavefunction is thus a superposition of both propagating and evanescent modes. Although the transmission is carried by the propagating modes only, the evanescent modes {\it indirectly} affect the transmission because  they contribute to realize the wavefunction matching Eq.(\ref{bc-single-interface}) at the interface. Effectively, one could  consider the two RSOC regions as ``leads" with (massless) propagating modes $\Psi^{}_{pr}$ with misaligned spins, which are coupled to (massive) evanescent modes $\Psi^{}_{ev}=\xi e^{\pm \kappa_{+} x}$ localized at the interface, whose spinor $\xi$ effectively performs the spin  rotation, thereby favoring the transmission\cite{gogin_2022}. Of course, in the case of equal RSOC signs, the misalignment of the propagating modes is less pronounced than in the case of opposite RSOC signs, and this is why for $|E|<E_Z$ the transmission is typically higher in Fig.\ref{Fig2}(a) than in \ref{Fig2}(b).
However, as the energy approaches (say) the upper edge of the magnetic gap,  $E\rightarrow  E_Z$, each evanescent mode   becomes the state related to the minimum of the upper band (see Fig.1): Its spin  gets locked along $x$ and its decay    lengthscale diverges.  This makes it effectively unable to guarantee a finite transmission, since a spatially uniform mode has a vanishing spatial overlap with a propagating mode.  The transmission thus vanishes with a cusp behavior because the evanescent mode wavevector~$ \kappa_{+}$ vanishes with an infinite slope as a function of energy $E \rightarrow E_Z$.
To a more quantitative level,  one can see that in such a limit $\Psi_{ev}$ can contribute to the first interface matching Eq.(\ref{bc-single-interface})  only with a spin along $x$, and its contribution to the spatial derivative   in the second Eq.(\ref{bc-single-interface}) vanishes. It turns out that the only way to realize a wavefunction matching of both Eqs.(\ref{bc-single-interface}) at the interface is to have a totally reflected mode, as one can verify by plotting the wavefunction profile (not shown here). 
A similar effect occurs at the lower magnetic gap edge $E\rightarrow  -E_Z$ when the region is in the Rashba-dominated regime, since the evanescent mode becomes the state related to the local maximum of the lower band (see Fig.\ref{Fig1}).

\section{The doubly gated Nanowire}
\label{sec-5}
The analysis of the single interface case carried out in the previous Section indicates that the conductance heavily depends on the relative sign between the two RSOC regions. However, a sharp separation between the two RSOC regions is in fact an idealization. In a realistic NW setup, where  the RSOC can be locally controlled by   different gates,  a finite distance $d$ separates the gates, as shown in the configurations sketched in Fig.\ref{Fig4}. In particular, panel (a) describes a realization with a NW deposited on a substrate and covered by two $\Omega$-gates\cite{kouwenhoven_2017a}.{In this kind of setup the NW is usually separated from the metallic gate by a thin (few nm) dielectric with high relative dielectric constant, which --at a given gate voltage-- enhances the interface field causing the RSOC\cite{sasaki_2017,sasaki-NTT-review}. 
In contrast, Fig.\ref{Fig4}(b) illustrates a suspended NW coupled to two pairs of side gates with the advantage of strongly reducing the presence of defects\cite{giazotto_2019}.  In both cases the gate voltages control the magnitude and the sign of the RSOC. 

We now want to take into account the finite size $d$ of the central region separating the two gated NW regions, which was neglected in the preliminary analysis of Sec.\ref{sec-4}.  
For definiteness we shall assume that the RSOC in the central region  is negligible, so that it is  in the strongly Zeeman-dominated regime, and we shall adopt the following profile
\begin{equation}\label{alpha-double}
\alpha(x)= \left\{ \begin{array}{lcl}  \alpha_{L}   & \mbox{for} & x <-d/2\\ & & \\ 0  & \mbox{for} & |x| <d/2 \\ & & \\ \alpha_{R}   & \mbox{for} & x >d/2 
\end{array} \right. \quad.
\end{equation}
Moreover, we shall set the RSOC   to be positive in the right region ($\alpha_R>0$), while for the left region we will consider  both positive and negative sign of $\alpha_L$.  
\begin{figure}[h!]
    \includegraphics[width=\linewidth]{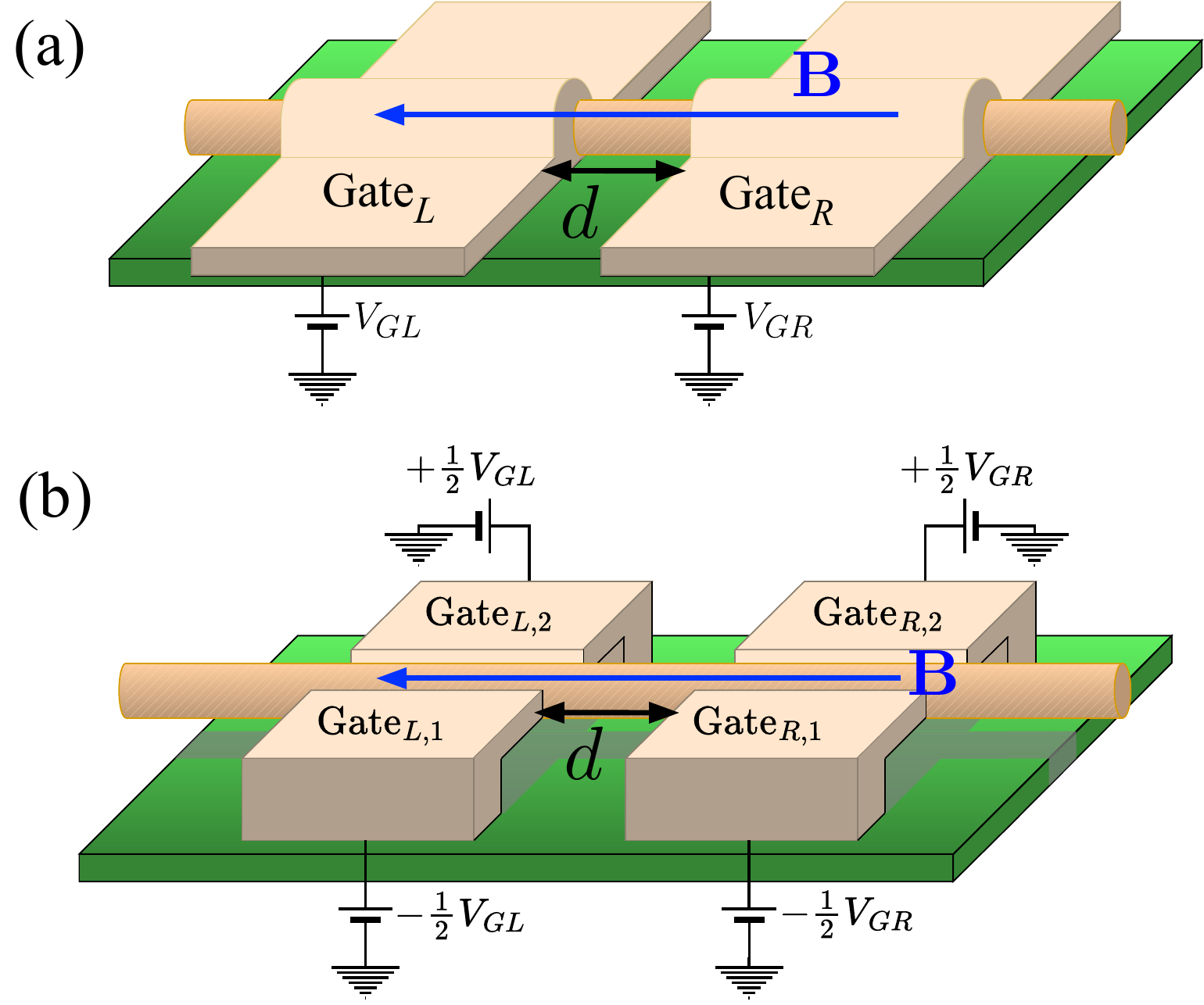}
  \caption{Two possible realizations of a doubly gated NW, where  two different NW regions, characterized by different RSOC, are separated by a distance $d$.  (a) The NW lies on a substrate and is contacted to two $\Omega$-gates (see e.g. Ref.[\onlinecite{kouwenhoven_2017a}]). (b) The NW is suspended and coupled to two pairs of side gates (see e.g. Ref.[\onlinecite{giazotto_2019}]). } 
  \label{Fig4}
\end{figure}

\begin{figure} 
    \includegraphics[width=\linewidth]{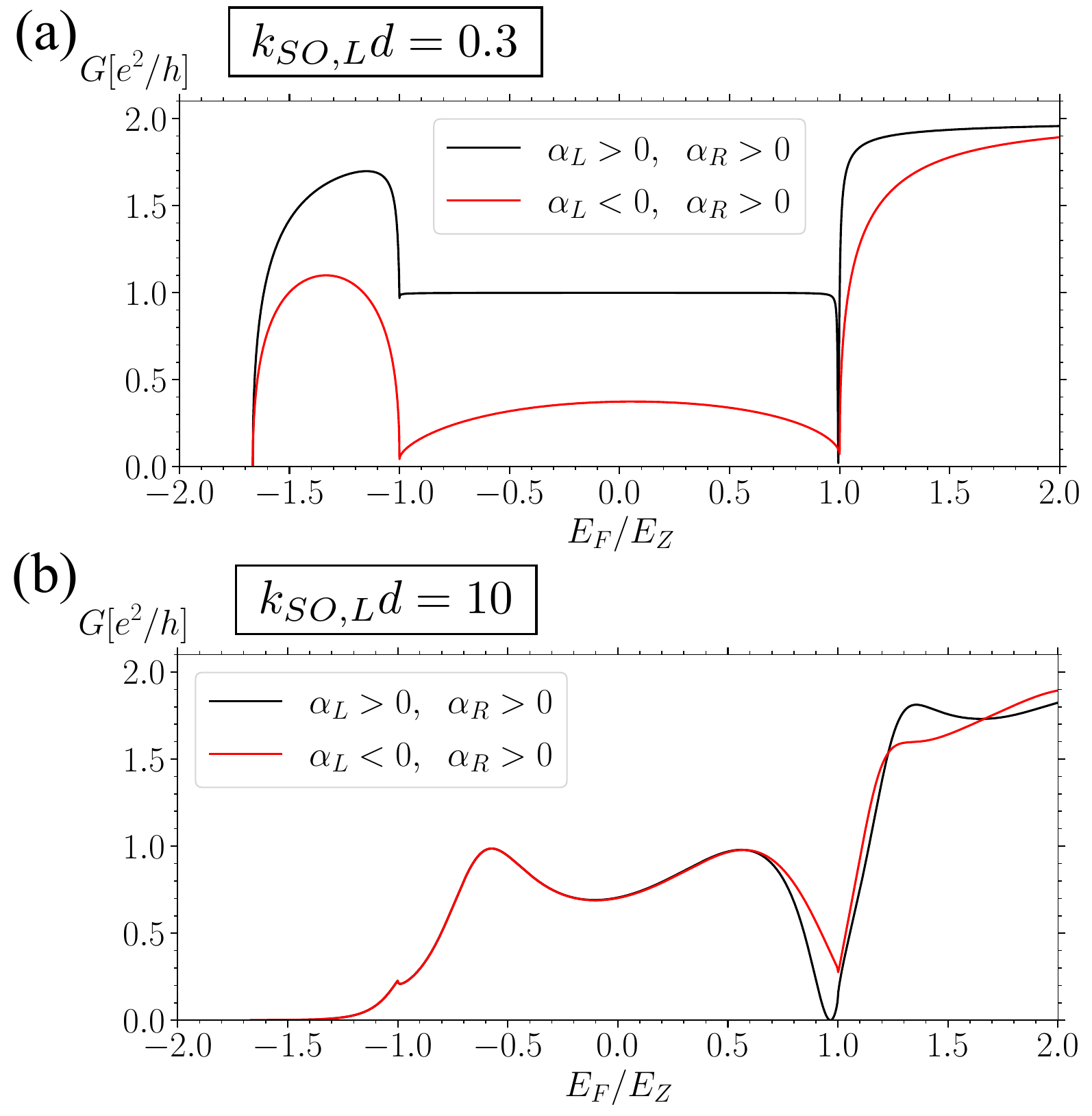}
    \caption{\small   The conductance of the double gate configuration [see Eq.(\ref{alpha-double})], where both gated regions are in the Rashba-dominated regime ($E_{SO,L}=3 E_Z$, $E_{SO,R}=1.5 E_Z$), is plotted in units of conductance quantum $G_0={\rm e}^2/h$ as function of the energy $E_F/E_Z$. The two panels correspond to a different  value $d$ of the distance between the two gated regions, expressed in units of the shortest spin-orbit lengthscale $k_{SO,L}^{-1}$, namely short distance (panel (a)),  and large distance (panel (b)).  The black and red  curves refer to the case where the two regions have equal and opposite sign of the RSOC, respectively. }
  \label{Fig5}
\end{figure}
 
In  Fig.\ref{Fig5} we  analyze the conductance $G$  as a function of the Fermi energy over the entire spectrum. We focus on the case where the two regions are both in the Rashba-dominated regime. The two spin-orbit energies Eq.(\ref{ESOj-def}) identify two  spin-orbit wavevectors 
\begin{equation}\label{kSOj-def}
k_{SO,j}=\frac{\sqrt{2 m^* E_{SO,j}}}{\hbar}=\frac{m^* |\alpha_j|}{\hbar^2} \hspace{1cm} j=L,R\quad,
\end{equation}
and the related spin-orbit lengths $k_{SO,j}^{-1}$. For definiteness we take  $E_{SO,L}=2 E_{SO,R}$ and  express the gate separation $d$ in units of the shorter spin-orbit length $k_{SO,L}^{-1}$.  In particular,  Fig.\ref{Fig5}(a) illustrates the result for a short gate separation: The conductance is practically perfect inside the magnetic gap when the two regions have equal RSOC sign (black curve), whereas it is suppressed when the two regions have opposite RSOC sign (red curve). Moreover, transmission is almost symmetric in energy and flat around the middle of the magnetic gap ($E=0$). The result  does not deviate much from the one obtained in Fig.\ref{Fig2} for the ideal $d\rightarrow 0$ limit. However, for large gate separation [see Fig.\ref{Fig5}(b)], the difference between the two cases of equal or opposite RSOC signs is strongly reduced, except for a small difference near the gap edge $E/E_Z = +1$.  
This is due to the fact that, for larger separation $d$, the electron spin has a spatial room to re-adapt to the different orientation imposed by the opposite RSOC sign.  
We also note the cusps with vanishing conductance discussed in the single interface problem are modified by the   finite distance $d$ between the gates. First, the effect of vanishing conductance may still be present, but it occurs at   values of  Fermi energy that can slightly differ from the magnetic gap boundaries, as  is   visible e.g. in the black curve of Fig.\ref{Fig5}(b), near the upper magnetic gap boundary $E_F=+ E_Z$. This makes the doubly gated NW configuration similar  to   the configuration of a NW contacted to leads\cite{sanchez_2008,aguado_2015}. Second,  the cusp behavior might occur at a finite conductance value. This can be the case even for a short distance $d$ [see Fig.\ref{Fig5}(a)] and  is  especially visible near  the lower magnetic gap boundary $E_F=- E_Z$. In this case, only for very short distance ($k_{SO,L} d < 0.01$) one recovers the vanishing conductance of the single interface problem. 
\begin{figure}
    \includegraphics[width=\linewidth]{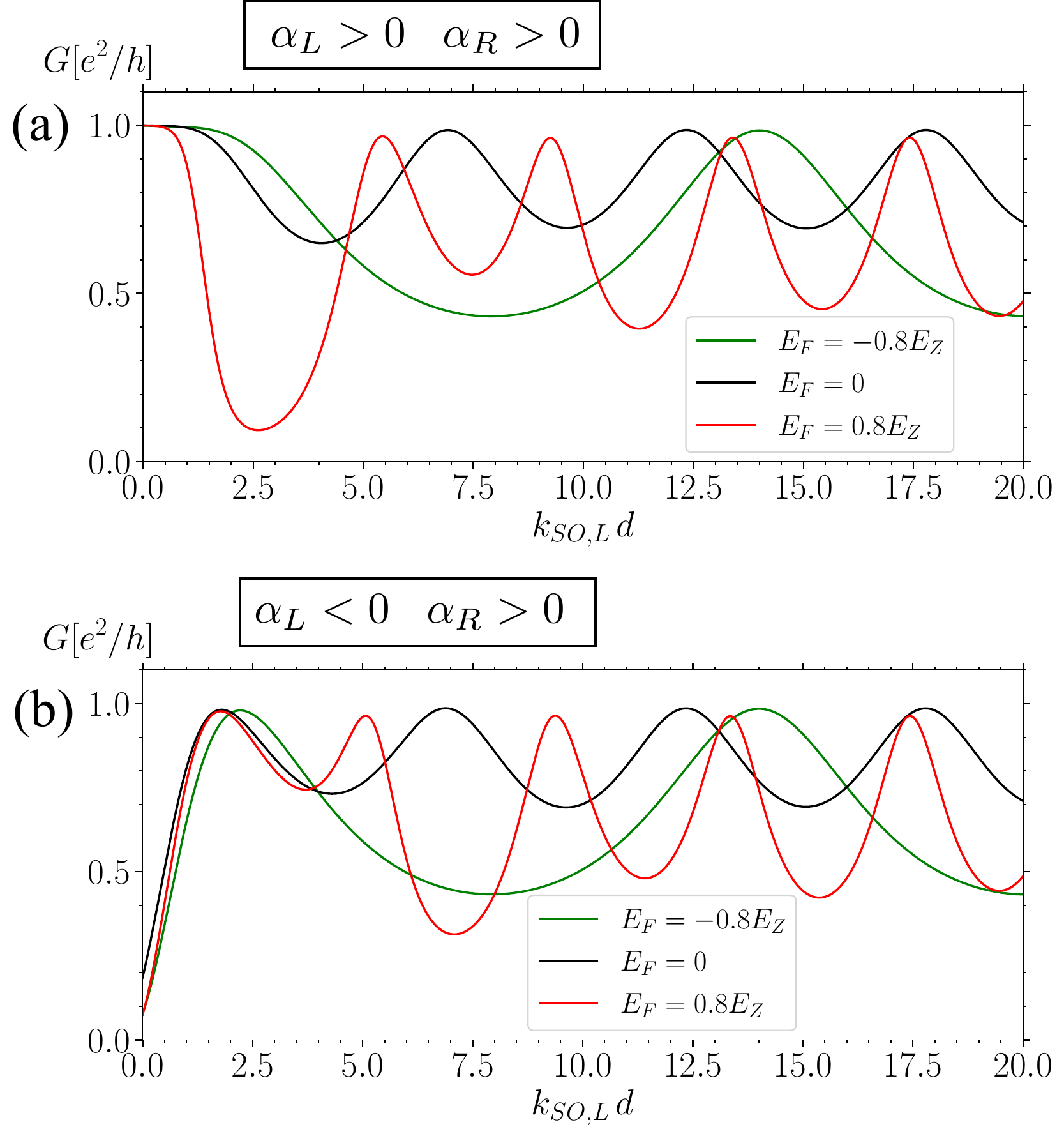}
  \caption{The double gate configuration, where both gated regions are in the Rashba-dominated regime ($E_{SO,L}=3 E_Z$, $E_{SO,R}=1.5 E_Z$). The conductance, evaluated at three fixed values of Fermi energy inside the gap, $E_F=-0.8 E_Z$ (green curve) $E_F=0$ (black curve) and $E_F=+0.8 E_Z$ (red curve), is plotted as function of the distance $d$ between the gates. Panels (a) and (b) illustrate the cases of equal and opposite RSOC sign,  respectively}
  \label{Fig6}
\end{figure}

Figure~\ref{Fig6} illustrates how the conductance, evaluated at three different Fermi energy values inside the magnetic gap, behaves as a function of  the distance $d$ between the two gates. At  short distance, $k_{SO,L}  d \ll1$, the   behavior of $G$ strongly depends on whether the signs of the two RSOCs are equal [panel (a)] or opposite [panel (b)]: In the former case it is  independent of the distance $d$  for all energy values ($G\cong G_0$), while in the latter case it  linearly grows with the distance and it is small ($G\ll G_0$). As the distance $d$ becomes of the order of the spin-orbit length, $k_{SO,L}  d\sim 1$, in both cases the conductance exhibits a crossover to an oscillatory behavior. This originates from the fact that at the boundaries of the central region part of the electron wave is reflected, giving rise to interference conditions that depend on the distance $d$. Because the electron wavelength depends on the Fermi energy, such oscillatory pattern depends on $E_F$ too, as one can see from the different period characterizing the various curves. Notably, for large distance $k_{SO,L}  d \gg1$, such oscillatory behavior  becomes independent of the relative RSOC signs of the two regions, as can be appreciated by comparing panels (a) and (b).  \\

{\it Implementation with InSb NWs}. Let us now consider a specific implementation of the setup of Fig.\ref{Fig4} with a InSb NW, with effective mass $m^*=0.015 m_e$  and $g$-factor $g\simeq 50$, exposed to a magnetic field corresponding to a Zeeman energy $E_Z=0.1 {\rm m eV}$. For definiteness,  for the RSOC  in the right region we have fixed a positive value $\alpha_R\simeq 0.552 \,{\rm eV \AA}>0$ corresponding to a spin-orbit energy $E_{SO,R}=0.3 \, {\rm m eV}=3 E_Z$, so that the right region is in the Rashba-dominated regime. Then,  by varying  $\alpha_L$  over a broad range of positive and negative values, we have analyzed how the conductance in the middle of the magnetic gap ($E_F=0$) depends on the ratio $\alpha_L/\alpha_R$.  The result is plotted in Fig.\ref{Fig7} for four different values of the distance $d$ between the two gates, which can be compared to the reference lengthscale given by the fixed spin-orbit length of the right-side is $k^{-1}_{SO,R}\simeq 92\, {\rm nm}$. As one can see, when $d$ is much shorter than the latter scale (black curve) the conductance $G$ exhibits a strongly asymmetric curve as a function of $\alpha_L/\alpha_R$. In this case $G$ depends only weakly on the magnitude of the RSOC, and strongly on the sign. Indeed a sign switch    of $\alpha_L$ from positive to negative changes the  transmission    from high to low values. In contrast, when $d$ becomes of the order of the spin-orbit length (red and green curves), the asymmetry of the curve gradually softens and eventually, for a large separation $d=1\,{\rm \mu m}$ (blue curve), the conductance exhibits a completely symmetric Lorentzian-like behavior as a function of $\alpha_L/\alpha_R$. In this regime only the magnitude of the RSOC determines the conductance, while  the relative sign of the RSOC plays no role.

\begin{figure}[h!]
    \includegraphics[width=\linewidth]{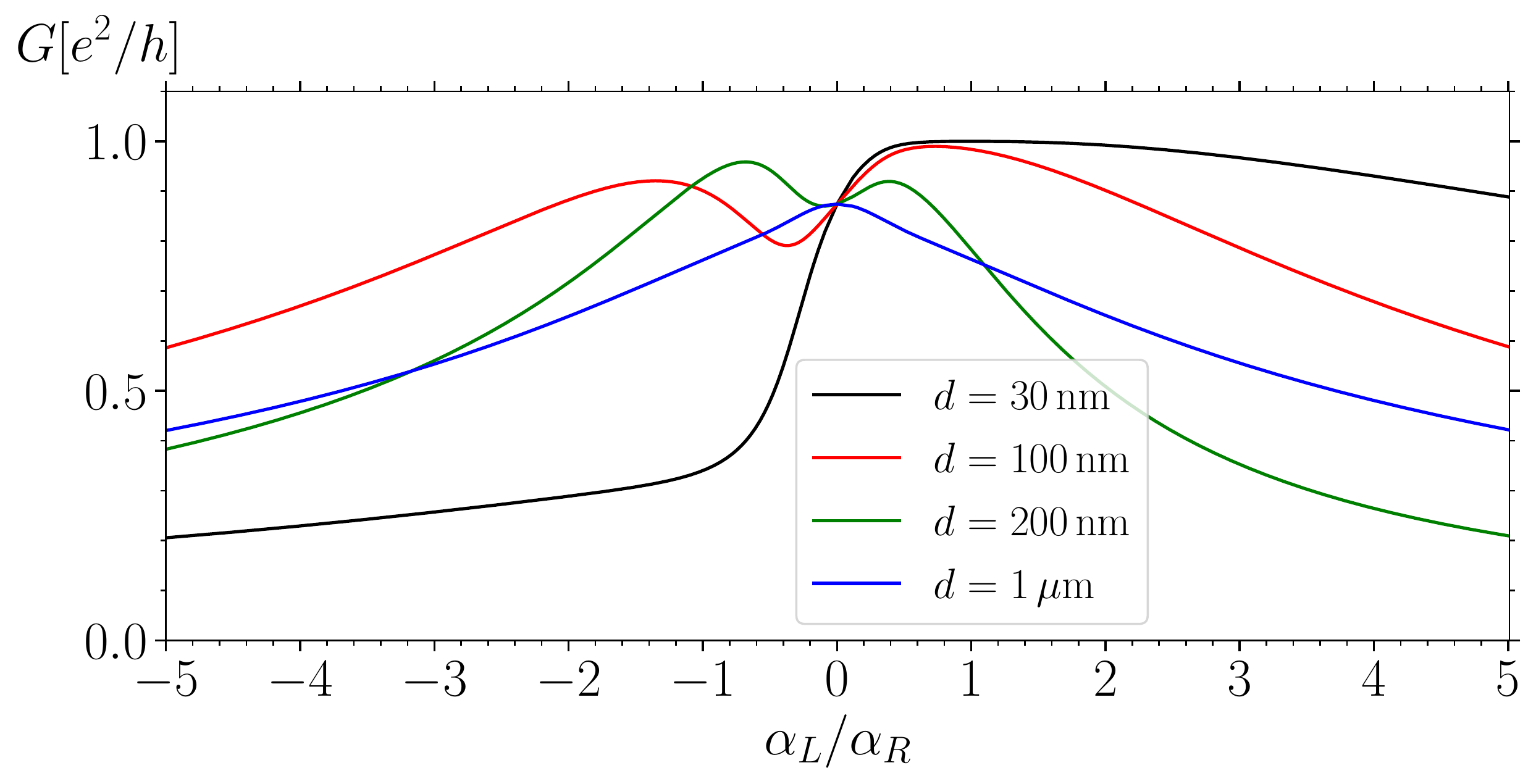}
    \caption{\small  The implementation of the double gate configuration with a InSb NW ($m^*=0.015 m_e$) exposed to a magnetic field with an Zeeman energy  $E_Z=0.1\, {\rm meV}$. The RSOC  in the right region is  fixed to be positive, $\alpha_R>0$, and corresponding to a spin-orbit energy $E_{SO,R}=0.3 \,{\rm meV}$ (Rashba dominated regime), corresponding to a spin-orbit length $k_{SO,R}^{-1} \simeq 92\,{\rm nm}$. The RSOC $\alpha_L$ in the left region is varied in magnitude and sign: The conductance at $E_F=0$ (middle of the magnetic gap) is shown as a function of the ratio $\alpha_L/\alpha_R$, for different values of the separation $d$ between the two gates. }
    \label{Fig7}
\end{figure}

{\it The helical limit.} We conclude this section by a comment about the helical limit. 
As is well known, in the strongly Rashba-dominated regime, $E_{SO} \gg E_Z$, the propagating states in the magnetic gap  are helical, i.e. they exhibit a locking between the propagation direction\cite{vonoppen_2010,dassarma_2010,streda,depicciotto_2010,loss_PRB_2011,lutchyn_2012,loss_PRB_2017}, similarly to the edge states of a 2D Topological Insulator. Importantly, in a NW the helicity is controlled by the sign of the RSOC. Explicitly, if $\alpha>0$ right-movers are characterized by spin-$\uparrow$ and left-movers  by spin-$\downarrow$, while the opposite occurs for $\alpha<0$, as is clear from the blue/red arrows in Fig.\ref{Fig1}(a). This opens up the possibility  to conceive transport configurations that would be quite hard in 2D Topological Insulators. Let us now consider  the inhomogeneous setup sketched in Fig.\ref{Fig4} in the limit where both gated regions are set to the strongly Rashba-dominated and exhibit opposite   RSOC signs. This  implements  a peculiar setup, namely a junction of helical states with opposite helicity, the so called Dirac paradox.   As is well known, for a NW with homogeneous RSOC, the low energy transport within the magnetic gap, i.e. in the energy range $|E_F|\ll E_Z$, is well captured by  a massless Dirac model  describing the helical states.   One may thus wonder whether in such energy window the behavior of the conductance for the inhomogeneous setup of Fig.\ref{Fig4} could be deduced from the low energy massless Dirac models. This is indeed the case when the RSOC takes the same sign in both gated regions: The low energy effective theory  is well captured by two  massless Dirac models with the same helicity, one on each region,  and one recovers a perfect transmission at low energy\cite{peres}. However, in the case of opposite RSOC signs, the customary massless Dirac models are not sufficient to describe the low-energy transport. Instead it is possible to show\cite{gogin_2022} that additional massive Dirac modes have to be included on each side to correctly account for the low-energy behavior at $|E_F|\ll E_Z$. Although in such  energy window  these modes are evanescent and carry no current directly, they do contribute to the current indirectly by favoring the correct wavefunction matching at the interface. The transmission thus  depends on the parameter $k_{SO} d$ in a non-monotonous way\cite{gogin_2022}, in agreement with the more general results presented here.

\section{Conclusions}
\label{sec-6}
In Conclusion, in this paper we have analyzed the effects of the sign of the RSOC  on electron transport in a one-dimensional electron channel in the quantum coherent regime. In the case of a NW with homogeneous RSOC and contacted to two electrodes, despite the spin orientation of the eigenstates   depends on the sign of the RSOC (see Fig.\ref{Fig1}),  we have shown  that  the transport properties are independent of $\mbox{sgn}(\alpha)$.   In fact, in Sec.\ref{sec-3}  we have identified two essential ingredients for the effects of the sign to be observed, namely the presence of i) at least two regions with different RSOC and ii) an additional Zeeman field that is non collinear with the effective RSOC magnetic field. In Sec.\ref{sec-4} we have first analyzed the electron  transport across an ideal interface separating two regions with different RSOC. We have shown that, when the sign of the RSOC is the same in both regions, the transmission is practically perfect, {\it regardless} of the regimes (Rashba- or Zeeman-dominated) of the two regions [see Fig.\ref{Fig2}(a)]. In this case the magnetic field essentially determines the number of conducting channels and the conductance can be tuned to integer values of the conductance quantum [see Fig.\ref{Fig3}(a)]. The scenario changes when the RSOC exhibits opposite signs at the two interface sides. In particular, while transmission remains extremely good when both regions are in the Zeeman regime, it reduces when one of the two regions enters the Rashba regime, and it gets suppressed when both regions are in the Rashba regime  [see Fig.\ref{Fig2}(b)]. In this case, the conductance can be continuously tuned  by varying the Zeeman field [see Fig.\ref{Fig3}(b)]. 

Then, in Sec.\ref{sec-5} we have considered a more realistic NW setup, where two different RSOC regions are realized by coupling the NW  to various gates (see Fig.\ref{Fig4}), and we have taken into account the   finite distance $d$ between the gates. The impact of the sign is quite different in the two regimes of small and large separation $d$   (see Fig.\ref{Fig5}). Indeed, by analyzing the conductance as a function of the separation $d$  (see Fig.\ref{Fig6}), we have found that in the regime of short separation the conductance strongly depends on the relative  RSOC signs, namely for equal signs one has $G \cong G_0$, roughly independently of $d$, while for opposite signs one has $G \ll G_0$ growing linearly with $d$. In contrast, as the separation becomes comparable with the spin-orbit length, the conductance exhibits a crossover to an oscillatory behavior, originating from the quantum interference of the electron backscattering of the ends of the separation region. In particular, for large separation ($k_{SO,L} d \gg 1$) such oscillatory behavior becomes independent of the relative RSOC signs.

Furthermore,  we have considered a specific implementation of the setup of Fig.\ref{Fig4} with InSb NWs, for four realistic values of separation $d$ between the gates. Focussing on the midgap conductance $G$ of the setup as a function of the ratio $\alpha_L/\alpha_R$ between the two RSOCs (see Fig.\ref{Fig7}), we have been able to identify the conditions   where $G$ is mainly determined by the sign or by the magnitude of the RSOC. In particular, for a short separation $d=30\,{\rm nm}$,  the conductance mainly depends on the relative RSOC signs, and one can tune $G$ from a small to a high fraction of ${\rm e^2}/h$ by varying the ratio $\alpha_L/\alpha_R$ from negative to positive values (black curve in Fig.\ref{Fig7}). In this regime, the RSOC sign thus represents an actual additional knob to switch on/off the  electron current through the NW. 
In contrast, for a large separation $d=1\,{\rm \mu m}$,  the conductance becomes insensitive to the  sign of  
$\alpha_L/\alpha_R$ (see blue curve in Fig.\ref{Fig7}) and is mainly dependent on the RSOC magnitude only. Finally, we have discussed that, for the particular range of small energies $|E_F|\ll E_Z$ inside the magnetic gap,  our findings are in agreement with the low-energy Dirac theory: While in the case of  two regions with equal RSOC signs the  low-energy electron transport  is well described by massless Dirac modes, in the case of opposite RSOC signs this is not the case, and additional massive Dirac modes have to be included.
 
A potentially interesting future development  of this work is to extend the analysis of electron transport through the doubly gated NW to the case of  {ac} gate voltages\cite{sherman_PRB_2013,malshukov_2003,malshukov_2005,dolcini_2012,loss_2016}, where two time-dependent RSOC are induced along the NW. Work is in progress along these lines.  
\acknowledgments
We are thankful to Lorenzo Rossi for fruitful discussions

\appendix
\begin{widetext}
\section{Classification of NW eigenstates}
\label{AppA}
As mentioned in the Main Text, the solution of the inhomogeneous RSOC problem is built by suitably  matching  the wavefunctions of each piecewise homogeneous NW region. To this purpose, for each energy value $E$, one has to classify all eigenstates (both  propagating  and evanescent) of a given NW region with RSOC $\alpha_{j}$. This Appendix is meant to provide this classification. In order to make the notation lighter, we shall use in this section $\alpha_{j}\rightarrow \alpha$ and $E_{SO,j}\rightarrow E_{SO}$.

{\it Propagating states.} Let us start from  the propagating modes $\psi(x)=\chi \exp[\pm i k x]$, where $\chi$ is a $2 \times 1$ spinor. Denoting $\varepsilon^0_k=\hbar^2 k^2/2m^*$ and inverting the energy spectra $E_{1,2}(k)=\varepsilon^0_k\mp \sqrt{E_Z^2+(\alpha k)^2}$ of the two bands ($\beta=1,2$) in favour of the energy $E$,   there are a priori four wavevectors for each value of $E$, namely two positive ones, $k_\pm(E)$, and two negative ones $-k_\pm(E)$,  where 
{\small
\begin{equation}\label{keta(E)-def}
     k_\eta(E;\alpha) = \frac{\sqrt{2 m^* }}{\hbar} \sqrt{E+2 E_{SO} +\eta \sqrt{4E E_{SO} + 4 E_{SO}^2 + E_Z^2}}   \hspace{2cm} \eta=\pm \quad.
\end{equation}
}

\noindent This can explicitly be seen from Fig.\ref{Fig8}(a)  for the Rashba-dominated regime  and in Figs.\ref{Fig9}(a) for the Zeeman-dominated regime, where $k_+$ and $k_{-}$ are distinguished by thick and thin curves, respectively.  
\noindent However, only real wavevectors ($k_\eta \in \mathbb{R}$) actually describe propagating modes and must be retained. For these wavevectors, the magnitude $v= \hbar^{-1} |\partial_k E|$ of the group velocity is expressed as a function of energy as
\begin{equation}\label{v-eta-def}
    v_\eta(E;\alpha)=\frac{\hbar k_\eta(E;\alpha)}{m^*} \frac{\sqrt{4 E E_{SO} + 4E_{SO}^2 +E_Z^2}}{\left|2 E_{SO}+\eta \sqrt{4 E E_{SO} + 4E_{SO}^2 +E_Z^2} \right|}  \hspace{2cm} \eta=\pm \quad.
\end{equation}

\begin{figure}
                 \includegraphics[width=14cm]{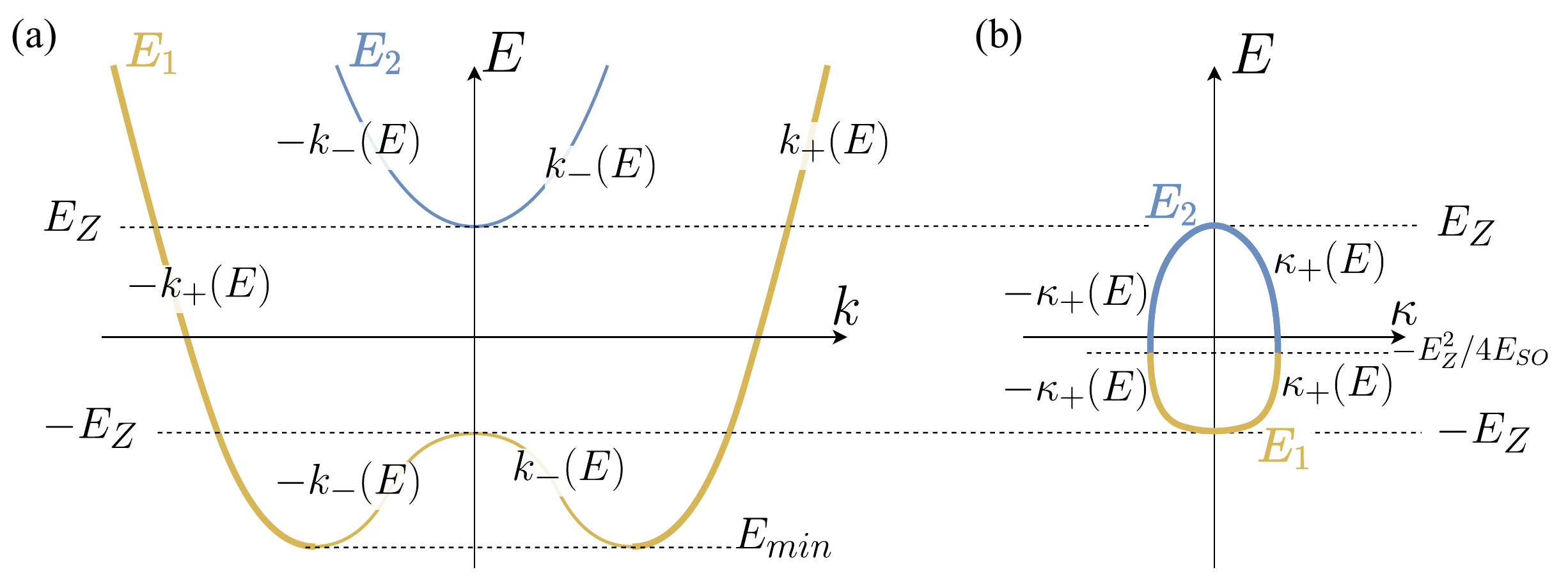}
    \caption{The Rashba dominated regime ($E_Z < 2 E_{SO}$).    (a) Spectrum of the   propagating modes, where thick and thin lines correspond to the the expressions $\pm k_{+}(E)$ and $\pm k_{-}(E)$ of the wavevector as a function of energy, respectively [see Eq.(\ref{keta(E)-def})]. Here   $E_{min}=-E_{SO} \left(1 + E_Z^2/4 E_{SO}^2  \right)$.   (b) Spectrum of the evanescent modes, where thick and thin lines correspond to $\pm \kappa_+(E)$ and to $\pm \kappa_-(E)$ [see Eq.(\ref{kappaeta(E)-def})], respectively. Note that panel (b)  can also be seen as the spectrum of panel (a) in the magnetic gap and along the imaginary axis $k=i\kappa$.}
    \label{Fig8}
\end{figure}

\noindent For each wavevector related to the energy $E$, the spinor~$\chi$ acquires a different expression depending on the band $\beta$. Explicitly, the two eigenvectors of the lower band ($\beta=1$) and the upper band  ($\beta=2$) are
\begin{eqnarray}
    \chi_1(\pm k_\eta(E;\alpha)) =\Vector{\cos\left(\frac{ \theta(\pm k_\eta(E;\alpha))}{2} \right)\\ \\ \sin\left(\frac{ \theta(\pm k_\eta(E;\alpha))}{2} \right)}\label{eq:PropagatingSpinor_E_band1}
\end{eqnarray}
\begin{eqnarray}
    \chi_2(\pm k_\eta(E;\alpha)) = \Vector{-\sin\left(\frac{\theta(\pm k_\eta(E;\alpha))}{2}  \right)\\ \\ \cos\left(\frac{\theta(\pm k_\eta(E;\alpha))}{2}  \right)}  \label{eq:PropagatingSpinor_E_band2}
\end{eqnarray}
respectively, where  

{\small
\begin{align}
    \cos\left( \frac{\theta(\pm k_\eta(E;\alpha)) }{2}  \right) &= \sqrt{\frac{1}{2}\left(1 \pm \sign(\alpha) \frac{\sqrt{4 E_{SO}(E + 2 E_{SO}) + \eta 4 E_{SO} \sqrt{4 E E_{SO} + 4 E_{SO}^2 + E_Z^2} }}{\left|\sqrt{4 E E_{SO} + 4 E_{SO}^2 + E_Z^2}+ \eta 2 E_{SO} \right|} \right)}\label{eq:half_theta1}\\ 
    \nonumber \\
    \sin\left( \frac{\theta(\pm k_\eta(E;\alpha))}{2}  \right) &= \sign(h_\perp) \sqrt{\frac{1}{2}\left(1 \mp \sign(\alpha) \frac{\sqrt{4 E_{SO}(E + 2 E_{SO}) + \eta 4 E_{SO} \sqrt{4 E E_{SO} + 4 E_{SO}^2 + E_Z^2} }}{ \left| \sqrt{4 E E_{SO} + 4 E_{SO}^2 + E_Z^2}+ \eta 2 E_{SO}  \right|} \right)}\label{eq:half_theta2}
\end{align}
}
\noindent The eigenstates (\ref{eq:PropagatingSpinor_E_band1})-(\ref{eq:PropagatingSpinor_E_band2}) identify a spin lying in the $x$-$z$ plane and forming an angle     $\theta$ with the $z$-axis. This enables one to obtain the pattern of spin-orientation in the two bands in Fig.\ref{Fig1}.\\

{\it Evanescent states.} One can proceed in a similar way for the evanescent modes $\psi(x)=\xi \exp[\pm \kappa x]$, where $\xi$ is a spinor. Their spectrum $E_{1,2}(\kappa)=-\varepsilon^0_\kappa\mp \sqrt{E_Z^2-(\alpha \kappa)^2}$  consists of two bands ($\beta=1,2$) touching at the energy $E_{min}=-E_Z^2/4 E_{SO}$ for $|\kappa|=E_Z/|\alpha|$, as depicted in Fig.\ref{Fig8}(b) for the Rashba-dominated regime  and in Figs.\ref{Fig9}(b) and (c) for the weak and strong Zeeman  regime, respectively. When inverting the spectrum 
 in favour of energy, one obtains
\begin{equation}
    \kappa_{\eta}(E;\alpha) = \frac{\sqrt{2 m^* }}{\hbar}\sqrt{-E-2 E_{SO} +\eta \sqrt{4E E_{SO} + 4 E_{SO}^2 + E_Z^2}} \hspace{2cm} \eta=\pm  \label{kappaeta(E)-def}
\end{equation}

\begin{figure} 
                \includegraphics[width=15cm]{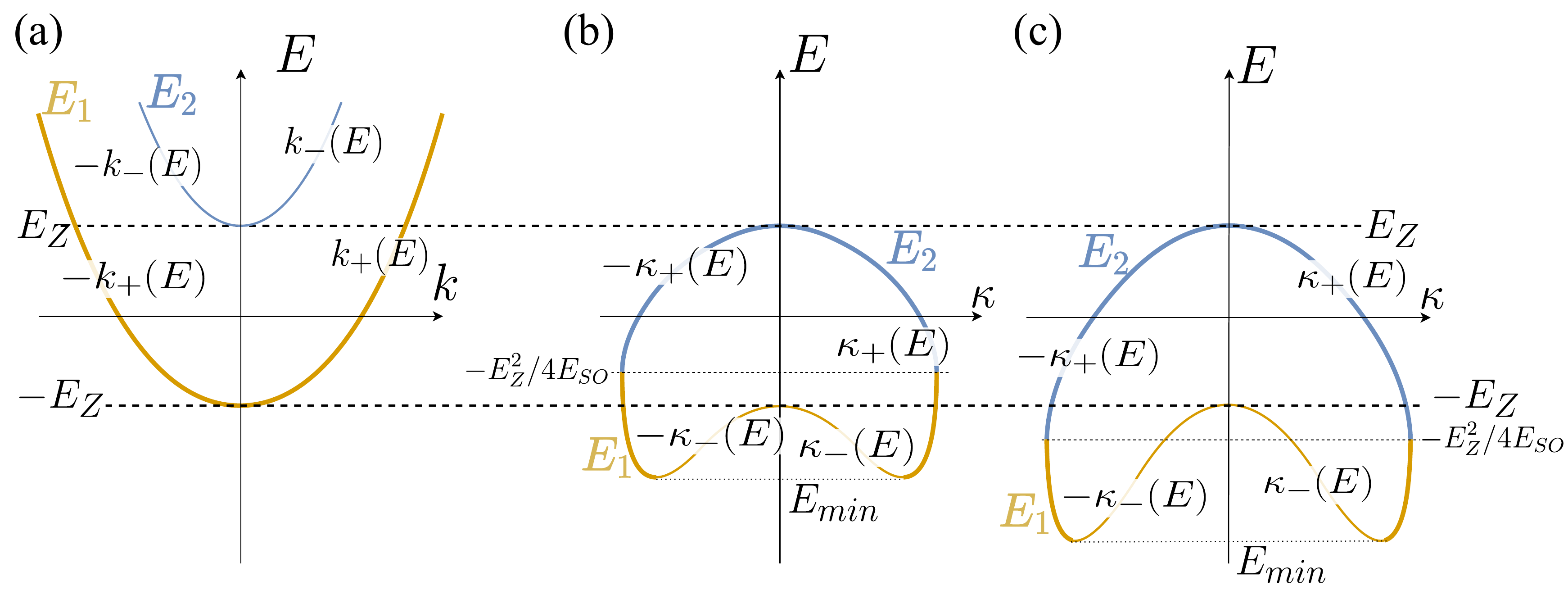}
    \caption{ The Zeeman dominated regime ($E_Z > 2 E_{SO}$). (a) The spectrum of the propagating modes. Thick and thin lines correspond to $\pm k_{+}(E)$ and $\pm k_{-}(E)$, respectively [see Eq.(\ref{keta(E)-def})]. (b) The spectrum of the evanescent modes in the weak Zeeman  regime $2 E_{SO} < E_Z < 4E_{SO}$; (c) The spectrum of the evanescent modes in the strong Zeeman   regime ($4 E_{SO} < E_Z$). Thick and thin lines correspond to $\pm \kappa_+(E)$ and to $\pm \kappa_-(E)$ [see Eq.(\ref{kappaeta(E)-def})].   In all plots  $E_{min} = -E_{SO} ( 1+ E_Z^2/4 E_{SO}^2)$.  }
    \label{Fig9}
\end{figure}
and the spinors acquire the form
\begin{align}
    \xi_1(\pm \kappa_\eta(E;\alpha)) &= \frac{1}{\sqrt{2}} \Vector{ e^{\mp i\, \arctan(\sinh(\theta( \kappa_\eta(E;\alpha))))}\\ 1} \label{eq:EvanescentSpinor_E1}\\
    \xi_2(\pm \kappa_\eta(E;\alpha)) &= \frac{1}{\sqrt{2}} \Vector{ - e^{ \pm i\, \arctan(\sinh(\theta( \kappa_\eta(E;\alpha))))}\\ 1} \label{eq:EvanescentSpinor_E2}
\end{align}
where
\begin{eqnarray}
  \sinh(\theta(\kappa_\eta(E;\alpha))) =  \sign(\alpha)   \, \frac{\sqrt{ -4 E_{SO}(E + 2 E_{SO}) + \eta \,4 E_{SO} \sqrt{4 E E_{SO} + 4 E_{SO}^2 + E_Z^2} }}{ \left|\sqrt{4 E E_{SO} + 4 E_{SO}^2 + E_Z^2} - \eta \,2 E_{SO} \right|   }
\end{eqnarray}
Note that, differently from the spinors of the propagating modes, the eigenstates (\ref{eq:EvanescentSpinor_E1})-(\ref{eq:EvanescentSpinor_E2}) identify a spin lying in the $x$-$y$ plane.

\begin{table} 
{\bf Rashba dominated regime ($ 2 E_{SO}>E_Z$) \,\,\,  ($E_{min}=-E_{SO} \left(1 + E_Z^2/4 E_{SO}^2  \right)$)} \\
{\small 
\begin{tabular}{|c||c|c||c|} 
    \hline
       {\bf energy range} 
    &  {\bf propagating modes}  
    &  {\bf velocity} 
    & {\bf evanescent modes} \\ 
    \hline
    $E_{min} < E < - E_Z  $ 
    & \begin{tabular}{@{}c@{}} 
        4 propagating states \\
        $\chi_1(\pm k_{-}(E)) e^{\pm i k_{-}(E) x}$ \\   
        $\chi_1(\pm k_{+}(E)) e^{\pm i k_{+}(E) x}$ \\ \\
        {\small [$k_\pm(E) \in E_1$ band]}
    \end{tabular} 
    & \begin{tabular}{@{}c@{}}
        \\
        $\mp v_{-}(E)$ \\   
        $\pm v_{+}(E)$ \\ \\ \\  
    \end{tabular}  
    & \mbox{\small No evanescent mode }   \\
    \hline
    $-E_Z <E<-\frac{E_Z^2}{4 E_{SO}}$ 
    & \begin{tabular}{@{}c@{}} 
        2 propagating states \\ 
        $\chi_1(\pm k_{+}(E)) e^{\pm i k_{+}(E) x}$\\ \\
        {\small [$k_+(E) \in E_1$ band]} 
    \end{tabular} 
    & \begin{tabular}{c} 
        \\
        $\pm v_{+}(E)$\\ \\ \\ 
    \end{tabular}  
    & \begin{tabular}{c}  
        $\xi_1(\pm \kappa_{+}(E)) e^{\pm \kappa_{+}(E)}$ \\ 
        \\
        {\small [$\kappa_+(E) \in E_1$ band]} \\  
    \end{tabular} \\ 
    \hline
    $-\frac{E_Z^2}{4 E_{SO}} <E<E_Z  $  
    & \begin{tabular}{@{}c@{}} 
        2 propagating states \\ 
        $\chi_1(\pm k_{+}(E)) e^{\pm i k_{+}(E) x}$\\  \\
        {\small [$k_+(E) \in E_1$ band]} 
    \end{tabular} 
    & \begin{tabular}{c}
        \\
        $\pm v_{+}(E)$\\ \\ \\ 
    \end{tabular}  
    & \begin{tabular}{c} 
        $\xi_2(\pm \kappa_{+}(E)) e^{\pm \kappa_{+}(E)}$ \\ 
        \\
        {\small [$\kappa_+(E) \in E_2$ band]} \\  
    \end{tabular} \\
    \hline
    $ E_Z<E$ 
    & \begin{tabular}{@{}c@{}} 4 propagating modes \\
        $\chi_2(\pm k_{-}(E)) e^{\pm i k_{-}(E) x}$ \\ 
        $\chi_1(\pm k_{+}(E)) e^{\pm i k_{+}(E) x}$ \\ \\
        {\small [$k_+(E) \in E_1$ band]} \\
        {\small [$k_-(E) \in E_2$ band]} \\ 
    \end{tabular} 
     &  \begin{tabular}{c}
        \\
        $\pm v_{-}(E)$ \\   
        $\pm v_{+}(E)$ \\ \\ \\    
    \end{tabular}  
    & \mbox{\small No evanescent mode }  \\
    \hline
\end{tabular} 
}
\caption{ \small 
    Eigenstates of the NW in the Rashba dominated regime ($2 E_{SO}>E_Z$).   For the propagating modes the wavevectors $k_\pm$ are given in Eqs.(\ref{keta(E)-def}), the  velocities    in Eq.(\ref{v-eta-def}) and the spinors $\chi_{1,2}$ in Eqs.(\ref{eq:PropagatingSpinor_E_band1})-(\ref{eq:PropagatingSpinor_E_band2}), while for the evanescent modes the wavevectors  $\kappa_\pm$ are given in Eq.(\ref{kappaeta(E)-def}) and the spinors $\xi_{1,2}$  are given in Eqs.(\ref{eq:EvanescentSpinor_E1}, \ref{eq:EvanescentSpinor_E2}).
}
\label{table-Rashba}
\end{table}
 
\begin{table}[h!]
{\bf Weak  Zeeman  regime ($2 E_{SO}<E_Z < 4 E_{SO}$)} \\
{\small
\begin{tabular}{|c||c|c||c|} 
    \hline
    {\bf energy range}     
    &  {\bf propagating modes}      
    &  {\bf velocity} & {\bf evanescent modes} \\ 
    \hline 
    $-E_{SO}\left( 1+\frac{E_Z^2}{4 E_{SO}^2}\right) <E< -E_Z $ 
    & \mbox{\small no propagating states} 
    &  
    & \begin{tabular}{@{}c@{}}
        $\xi_1(\pm \kappa_{+}(E)) e^{\pm \kappa_{+}(E) x}$ \\ 
        $\xi_1(\pm \kappa_{-}(E)) e^{\pm \kappa_{-}(E) x}$\\ \\
        {\small [$\kappa_\pm(E) \in E_1$ band]} 
    \end{tabular}\\
    \hline
    $-E_Z <E<-\frac{E_Z^2}{4 E_{SO}}$
    & \begin{tabular}{@{}c@{}} 
        2 propagating states \\ 
        $\chi_1(\pm k_{+}(E)) e^{\pm i k_{+}(E) x}$\\ \\
        {\small [$k_+(E) \in E_1$ band]} 
    \end{tabular} 
    & \begin{tabular}{@{}c@{}} 
        \\
        $\pm v_{+}(E)$ \\   
        \\ \\       
    \end{tabular}  
    & \begin{tabular}{@{}c@{}} 
        $\xi_1(\pm \kappa_+(E)) e^{\pm \kappa_+(E)}$\\ \\
        {\small [$\kappa_+(E) \in E_1$ band]} 
    \end{tabular}\\ 
    \hline
    $-\frac{E_Z^2}{4 E_{SO}} <E<E_Z  $ 
    & \begin{tabular}{@{}c@{}} 
        2 propagating states \\ 
        $\chi_1(\pm k_{+}(E)) e^{\pm i k_{+}(E) x}$\\ \\
        {\small [$k_+(E) \in E_1$ band]} 
    \end{tabular} 
    & \begin{tabular}{@{}c@{}} 
        \\
        $\pm v_{+}(E)$ \\   
        \\ \\       
    \end{tabular} 
    & \begin{tabular}{@{}c@{}} 
        $\xi_2(\pm \kappa_+(E)) e^{\pm \kappa_+(E)}$\\ \\
        {\small [$\kappa_+(E) \in E_2$ band]} 
    \end{tabular}\\ 
    \hline
    $E >  E_Z$ 
    & \begin{tabular}{@{}c@{}} 
        4 propagating states \\ 
        $\chi_2(\pm k_{-}(E)) e^{\pm i k_{-}(E) x}$ \\ 
        $\chi_1(\pm k_{+}(E)) e^{\pm i k_{+}(E) x}$\\ \\
        {\small [$k_-(E) \in E_2$ band]}\\
        {\small [$k_+(E) \in E_1$ band]}
    \end{tabular} 
    & \begin{tabular}{@{}c@{}} 
        \\
        $\pm v_{-}(E)$ \\   
        $\pm v_{+}(E)$\\ \\ \\ \\
    \end{tabular}  
    & \mbox{\small no evanescent states} \\
    \hline
\end{tabular}\\
}
\caption{ \small     Eigenstates of the NW in the  weak Zeeman   regime  ($2 E_{SO}<E_Z < 4 E_{SO}$). The meaning of symbols is the same as in Table \ref{table-Rashba}. }
\label{table-Weak-Zeeman}
\end{table}

\begin{table}[h!]
{\bf Strong  Zeeman  regime ($E_Z>4 E_{SO}$)}\\ 
{\small
\begin{tabular}{|c||c|c||c|} 
    \hline
    {\bf energy range}     
    &  {\bf propagating modes}      
    &  {\bf velocity} & {\bf evanescent modes} \\ 
    \hline 
    $-E_{SO}\left( 1+\frac{E_Z^2}{4 E_{SO}^2}\right) <E< -\frac{E_Z^2}{4 E_{SO}}$ 
    & \mbox{\small no propagating mode} 
    &  
    &\begin{tabular}{@{}c@{}}
        $\xi_1(\pm \kappa_{-}(E)) e^{\pm \kappa_{-}(E) x}$\\
        $\xi_1(\pm \kappa_{+}(E)) e^{\pm \kappa_{+}(E) x}$ \\ \\
        {\small [$\kappa_\pm(E) \in E_1$ band]}
    \end{tabular} \\
    \hline
    $-\frac{E_Z^2}{4 E_{SO}} <E< -E_Z   $ 
    & \mbox{\small no propagating states} 
    & 
    &\begin{tabular}{@{}c@{}}
        $\xi_1(\pm \kappa_{-}(E)) e^{\pm \kappa_{-}(E) x}$ \\ 
        $\xi_2(\pm \kappa_{+}(E)) e^{\pm \kappa_{+}(E) x}$\\ \\
        {\small [$\kappa_-(E) \in E_1$ band]}\\
        {\small [$\kappa_+(E) \in E_2$ band]}
    \end{tabular} \\\hline
    $-E_Z <E<E_Z  $ 
    & \begin{tabular}{@{}c@{}} 
        2 propagating states \\ 
        $\chi_1(\pm k_{+}(E)) e^{\pm i k_{+}(E) x}$\\ \\
        {\small [$k_+(E) \in E_1$ band]} 
    \end{tabular} 
    & \begin{tabular}{@{}c@{}} 
        \\
        $\pm v_{+}(E)$ \\   
        \\ \\       
    \end{tabular}
    & \begin{tabular}{@{}c@{}} 
        $\xi_2(\pm \kappa_+(E)) e^{\pm \kappa_+(E)}$\\ \\
        {\small [$\kappa_+(E) \in E_2$ band]} 
    \end{tabular}\\
    \hline
    $E >  E_Z$ 
    & \begin{tabular}{@{}c@{}} 
        4 propagating states \\ 
        $\chi_2(\pm k_{-}(E)) e^{\pm i k_{-}(E) x}$ \\ 
        $\chi_1(\pm k_{+}(E)) e^{\pm i k_{+}(E) x}$\\ \\
        {\small [$k_-(E) \in E_2$ band]}\\
        {\small [$k_+(E) \in E_1$ band]}
    \end{tabular}  
    & \begin{tabular}{@{}c@{}} 
        \\
        $\pm v_{-}(E)$ \\   
        $\pm v_{+}(E)$\\ \\ \\ \\
    \end{tabular}   
    & \mbox{\small no evanescent states} \\
    \hline
\end{tabular}
}
\caption{ \small 
    Eigenstates of the NW in the strong Zeeman  regime  ($E_Z > 4 E_{SO}$). The meaning of symbols is the same as in Table \ref{table-Rashba}.
}
\label{table-Strong-Zeeman}
\end{table}

{\it Three possible regimes.} Depending on the local value of $E_{SO}$ and $E_Z$, the NW region can be in three possible regimes, namely Rashba-dominated ($2E_{SO}>E_Z$), weak Zeeman regime ($2E_{SO}< E_Z<4 E_{SO}$) and strong Zeeman regime ($E_Z>4E_{SO}$). For each of these regimes, there are various energy ranges where the eigenstates acquire different expressions.
In Table \ref{table-Rashba} we have summarized  the result for the Rashba-dominated regime.
In particular,  propagating states exist for $E>E_{min}$, as can be see from  Fig.\ref{Fig8}(a). Their wavevectors   are given by Eq.(\ref{keta(E)-def}), the spinors $\chi_1$ and $\chi_2$, related to the lower and upper band, respectively, are given in Eqs.(\ref{eq:PropagatingSpinor_E_band1})-(\ref{eq:PropagatingSpinor_E_band2}), while the velocities are given in Eq.(\ref{v-eta-def}). In contrast, evanescent modes exist for $|E|<E_Z$, as shown in Fig.\ref{Fig9}(a), their wavevectors are given in Eqs.(\ref{kappaeta(E)-def}), and the related spinors, related to the upper and lower band, are given in Eqs.(\ref{eq:EvanescentSpinor_E1})-(\ref{eq:EvanescentSpinor_E2}).\\
As far as  the Zeeman-dominated regime is concerned ($E_Z>2 E_{SO}$), propagating modes exist for $E>-E_Z$ (see Fig.\ref{Fig8}(b)), while evanescent modes exist for $E_{min}<E<E_Z$. However,  the explicit expression and type (upper/lower band) of the NW states also depends whether the NW is  in the weak Zeeman subregime ($2 E_{SO}<E_Z<4E_{SO}$) or in the strong Zeeman subregime ($4E_{SO}<E_Z$), as shown in Figs.\ref{Fig9}(b) and (c).  The  expressions  of eigenstates in the weak and strong Zeeman regimes are summarized  in  Tables \ref{table-Weak-Zeeman} and \ref{table-Strong-Zeeman}, respectively.

\section{Details about the evaluation of the Scattering Matrix and the  Transmission coefficient}
\label{AppB}
Here we provide the technical details concerning the determination of the transmission coefficient for the system with  inhomogeneous RSOC setup, whence the conductance is straightforwardly determined from Eq.(\ref{conductance}).
To this purpose, one first has  compute the Transfer Matrix. We  exploit the fact that, since the system is mesoscopic, energy is conserved, and we can write the general solution as a superposition over the energy $E$ of stationary solutions $\hat{\Psi}_E(x,t)=\hat{\Psi}_E(x) \, e^{-i Et/\hbar }$. In turn, in the $j$-th region ($j=L,R=0,1$) the field $\hat{\Psi}_E(x)$  is expressed as a superposition of all possible modes (propagating and evanescent) characterizing the region at a fixed energy $E$.
These modes can be selected from Tables \ref{table-Rashba}, \ref{table-Weak-Zeeman} and \ref{table-Strong-Zeeman}, depending on the parameter regime (Rashba-dominated, weakly or strongly Zeeman dominated) and the energy sub-range.  

{\it The single interface case.} As an example, let us consider the case of one single interface at $x=0$, and assume that the left side is in the Rashba-dominated regime, while the right side is in the strongly Zeeman-dominated regime. Moreover, let us consider an energy value  inside the magnetic gap, $|E|<E_Z$. As can be seen from Tables \ref{table-Rashba}   and \ref{table-Strong-Zeeman},  in this energy range there are two propagating modes and two evanescent modes on each side of the interface.
Then, a scattering state  has the form
\begin{equation}\label{Ansatz-single-1}
 \hat{\Psi}_{E}^{}(x)=\frac{1}{\sqrt{2 \pi \hbar}}  \left\{ 
\begin{array}{lcl}
\begin{array}{l}
\frac{1}{\sqrt{v_{+}(E;\alpha_L)}} 
    \bigg( \hat{a}_{1E}^{(L)}  \chi_{1}(k_+(E;\alpha_L))\,e^{i k_+(E;\alpha_L) x} 
    + \hat{b}_{1E}^{(L)}  \chi_{1}(-k_+(E;\alpha_L))\, e^{-i k_+(E;\alpha_L) x} \bigg) +  \\ 
  \hspace{.5cm}  + \hat{f}_{\beta_E E}^{(L)} \ \xi_{\beta_E} (\kappa_+(E;\alpha_L)) e^{\kappa_+(E;\alpha_L)\, x}  
\end{array} 
& & x<0  \\ & & \\
\begin{array}{l}
\frac{1}{\sqrt{v_{+}(E;\alpha_R)}} \, 
  \bigg(   \hat{b}_{1E}^{(R)}  \chi_{1}(k_+(E;\alpha_R)) \,e^{i k_+(E;\alpha_R) x} + \hat{a}_{1E}^{(R)}  \chi_{1}(-k_+(E;\alpha_R)) \,e^{-i k_+(E;\alpha_R) x}  \bigg)+\\
    \hspace{.5cm} +   \hat{g}_{2 E}^{(R)} \ \xi_{2} (-\kappa_+(E;\alpha_R)) e^{-\kappa_+(E;\alpha_R)\, x}  
\end{array}& & x>0  
\end{array}
\right.
\end{equation} 
where  $\hat{a}_{1 E}^{(L/R)}$  denotes the fermionic operator related to the propagating mode incoming from the left/right side (originating from the lower band  $\beta=1$), $\hat{b}_{1E}^{(L/R)}$ describes  the one related to the propagating mode outgoing to the left/right (also originating from the lower band  $\beta=1$),  with $\chi_1$ being the related spinor given in Eq.(\ref{eq:PropagatingSpinor_E_band1}). Similarly
  $\hat{f}_{\beta_E E}^{(L)}$ and $\hat{g}_{2 E}^{(R)}$ are the evanescent mode operators and $\xi_1,\xi_2$ denote the related spinors  given in Eqs.(\ref{eq:EvanescentSpinor_E1})-(\ref{eq:EvanescentSpinor_E2}).  For the Rashba-dominated side on the left,  from Table \ref{table-Rashba} one sees that  the evanescent mode originates   from the  lower complex band ($\beta_E = 1$) for  $E < - E_Z^2/4 E_{SO}$ and  from the upper complex  band  ($\beta_E = 2$) for  $E > - E_Z^2/4 E_{SO}$, whereas for the strongly Zeeman-dominated side on the right we deduce from Table \ref{table-Strong-Zeeman} that the evanescent mode originates  from the  upper complex  band $(\beta=2$) on the Zeeman-side.
Inserting Eq.(\ref{Ansatz-single-1}) into the boundary conditions~(\ref{bc-single-interface}) at the interface $x=0$, the latter can be rewritten in a matrix form as
\begin{equation}
    M_E^{(L)}(0)   \begin{pmatrix}   \hat{a}_{1E}^{(L)}  \\ \hat{b}_{1E}^{(L)} \\ \hat{f}_{\beta_E E}^{(L)} \\ 0  \end{pmatrix}= M_E^{(R)}(0) 
    \begin{pmatrix}  \hat{b}_{1 E}^{(R)} \\  \hat{a}_{1E}^{(R)}  \\ 0 \\ \hat{g}_{2 E}^{(R)} \end{pmatrix}
    \label{eq:boundaryMatchingEq}
\end{equation}
where $M^{(L/R)}_E(0) \in \mathbb{C}^{4 \times 4}$ are two $4 \times 4$ complex boundary matrices at energy $E$ given by
{\small \begin{equation}
M_{E}^{(L)}(0) = 
\begin{pmatrix}
\frac{\chi_{1}(k_+)}{\sqrt{v_{+}}} &
\frac{\chi_{1}( - k_+)}{\sqrt{v_{+}}} &
\xi_{2} (\kappa_+)&
\xi_{2} (-\kappa_+)\\
&&&\\
i  \frac{( k_+ - \sigma_L k_{SO,L}  ) \ \chi_{1}(k_+)}{\sqrt{v_{+}}} &
- i  \frac{( k_+ + \sigma_L k_{SO,L} ) \ \chi_{1}( - k_+)}{\sqrt{v_{+}}} &
(\kappa_+ - i \sigma_L k_{SO,L} ) \xi_{2} (\kappa_+)&
- (\kappa_+ + i \sigma_L k_{SO,L} ) \xi_{2} (-\kappa_+)
\end{pmatrix}
 \label{ML-ris}
\end{equation}
}
and
{\small 
\begin{equation}
M_{E}^{(R)}(0) = 
\begin{pmatrix}
\frac{\chi_{1}(k_+)}{\sqrt{v_{+}}} &
\frac{\chi_{1}( - k_+)}{\sqrt{v_{+}}} &
\xi_{\beta_E} (\kappa_+)&
\xi_{\beta_E} (-\kappa_+)\\
&&&\\
i  \frac{( k_+ - \sigma_R k_{SO,R}) \ \chi_{1}(k_+)}{\sqrt{v_{+}}} &
- i  \frac{( k_+ + \sigma_R k_{SO,R}) \ \chi_{1}( - k_+)}{\sqrt{v_{+}}} &
(\kappa_+ - i \sigma_R k_{SO,R}) \xi_{\beta_E} (\kappa_+)&
- (\kappa_+ + i \sigma_R k_{SO,R}) \xi_{\beta_E} (-\kappa_+)
\end{pmatrix}
\label{MR-ris}
\end{equation}
 }
where $\chi_1$, $\xi_1$ and $\xi_2$ are the $2 \times 1$ spinors    in  Eq.(\ref{eq:PropagatingSpinor_E_band1}), (\ref{eq:EvanescentSpinor_E1})-(\ref{eq:EvanescentSpinor_E2}), respectively, and the dependence of   $\kappa_+$ and $v_{+}$ on $\alpha_L$ and $E$ [in Eq.(\ref{ML-ris})] and on  $\alpha_R$ and $E$ [in Eq.(\ref{MR-ris})] has been omitted for simplicity. 
Moreover, we have denoted $\sigma_{R,L}=\mbox{sgn}(\alpha_{R,L})$ and we have used the fact that the second Eq.(\ref{bc-single-interface}) at the interface $x=0$ can be rewritten as
\begin{equation}
\partial_x \hat{\Psi}(0^-)  = \partial_x \hat{\Psi}(0^+) -  i \left(\sigma_{R}  k_{SO,R}-\sigma_{L} k_{SO,L}\right) \sigma_z  \hat{\Psi}(0)
\end{equation}
 in terms of the spin-orbit wavevectors (\ref{kSOj-def}).  
From Eq.(\ref{eq:boundaryMatchingEq}), the transfer matrix relating the operators on the right side of interface   to  the operators on the left side,
\begin{equation}
   \begin{pmatrix}  \hat{b}_{1 E}^{(R)} \\  \hat{a}_{1E}^{(R)}  \\ 0 \\ \hat{g}_{2 E}^{(R)} \end{pmatrix} = W_E     \begin{pmatrix}  \hat{a}_{1E}^{(L)}  \\ \hat{b}_{1E}^{(L)} \\ \hat{f}_{\beta_E E}^{(L)} \\ 0 \end{pmatrix}  \quad,  \hspace{2cm} |E|<E_Z \label{Transfer-single-interface-1}
\end{equation}
is straightforwardly found to be $W_E  =  [ M_E^{(R)}(0)]^{-1} M_E^{(L)}(0)$. 
Since   Eqs.(\ref{Transfer-single-interface-1}) represent  four constraints for the six unknown operators, one can express four of them in terms   of the incoming operators $ \hat{a}_{1E}^{(L)}$ and $ \hat{a}_{1E}^{(R)}$. In particular, one can thus compute 
  the Scattering Matrix $S_E$, which relates the outgoing propagating modes to the incoming ones
  \begin{equation}\label{S-matrix-single-1}
    \Vector{\hat{b}_{1 E}^{(L)}\vspace{3pt} \\ \hat{b}_{1 E}^{(R)}}
    = \underbrace{\begin{pmatrix} r_E & t^\prime_E \\ & \\ t_E & r^\prime_E \end{pmatrix}}_{S_E} \Vector{\hat{a}_{1  E}^{(L)}\vspace{3pt} \\ \hat{a}_{1 E}^{(R)}}
\end{equation}
The transmission coefficient is computed as $T(E)=|t_E|^2=|t_E^\prime|^2$ and the linear conductance in the considered range $|E_F|<E_Z$ is obtained from Eq.(\ref{conductance}). \\

In a similar way, one can find the transmission in the energy range  $E>E_Z$ above the gap. In such a case, on each side of the interface there are two propagating modes and no evanescent mode on each side, so that the Scattering state solution now acquires the form
\begin{equation}\label{Ansatz-single-2}
 \hat{\Psi}_{E}^{}(x)=\frac{1}{\sqrt{2 \pi \hbar}}  \left\{ 
\begin{array}{lcl}
\begin{array}{l}
\frac{1}{\sqrt{v_{+}(E;\alpha_L)}} 
    \bigg( \hat{a}_{1E}^{(L)}  \chi_{1}(k_+(E;\alpha_L))\,e^{i k_+(E;\alpha_L) x} 
    + \hat{b}_{1E}^{(L)}  \chi_{1}(-k_+(E;\alpha_L))\, e^{-i k_+(E;\alpha_L) x} \bigg) +  \\ 
  \hspace{.5cm} + \frac{1}{\sqrt{v_{-}(E;\alpha_L)}} 
    \bigg( \hat{a}_{2E}^{(L)}  \chi_{2}(k_-(E;\alpha_L))\,e^{i k_-(E;\alpha_L) x} 
    + \hat{b}_{2E}^{(L)}  \chi_{2}(-k_-(E;\alpha_L))\, e^{-i k_-(E;\alpha_L) x} \bigg)  
\end{array} 
& & x<0  \\ & & \\
\begin{array}{l}
\frac{1}{\sqrt{v_{+}(E;\alpha_R)}} \, 
  \bigg(   \hat{b}_{1E}^{(R)}  \chi_{1}(k_+(E;\alpha_R)) \,e^{i k_+(E;\alpha_R) x} + \hat{a}_{1E}^{(R)}  \chi_{1}(-k_+(E;\alpha_R)) \,e^{-i k_+(E;\alpha_R) x}  \bigg)+\\
    \hspace{.5cm} +    \frac{1}{\sqrt{v_{-}(E;\alpha_R)}} 
    \bigg( \hat{b}_{2E}^{(R)}  \chi_{2}(k_-(E;\alpha_R))\,e^{i k_-(E;\alpha_R) x} 
    + \hat{a}_{2E}^{(R)}  \chi_{2}(-k_-(E;\alpha_R))\, e^{-i k_-(E;\alpha_R) x} \bigg) 
\end{array}& & x>0  
\end{array}
\right.
\end{equation} 
Inserting Eq.(\ref{Ansatz-single-2}) into the boundary conditions (\ref{bc-single-interface}) and proceeding as outlined above, one can find the Transfer Matrix
\begin{equation}
    \Vector{\hat{a}_{1 E}^{(R)}\vspace{3pt} \\ \hat{b}_{1  E}^{(R)}\vspace{3pt} \\ \hat{a}_{2  E}^{(R)}\vspace{3pt} \\ \hat{b}_{2  E}^{(R)}} 
    = W_E 
    \Vector{\hat{a}_{1  E}^{(L)}\vspace{3pt} \\ \hat{b}_{1  E}^{(L)}\vspace{3pt} \\ \hat{a}_{2  E}^{(L)}\vspace{3pt} \\ \hat{b}_{2 E}^{(L)}} \label{Transfer-single-interface-2}
\end{equation}
whence  the Scattering matrix $S_E$, which is now a $4 \times 4$ matrix, can straightforwardly be obtained
\begin{equation}\label{S-matrix-single-2}
\begin{pmatrix} \hat{b}_{1 E}^{(L)} \vspace{3pt} \\ \hat{b}_{2 E}^{(L)} \vspace{3pt} \\ \hat{b}_{1 E}^{(R)} \vspace{3pt} \\ \hat{b}_{2 E}^{(R)}\end{pmatrix}
    = \underbrace{\begin{pmatrix} \mathsf{r}_E &  \mathsf{t}^\prime_E \\ & \\  \mathsf{t}_E &  \mathsf{r}^\prime_E \end{pmatrix}}_{S_E} \begin{pmatrix} \hat{a}_{1 E}^{(L)} \vspace{3pt} \\ \hat{a}_{2 E}^{(L)}\vspace{3pt} \\ \hat{a}_{1 E}^{(R)}\vspace{3pt} \\ \hat{a}_{2 E}^{(R)}\end{pmatrix}
\end{equation}
with $\mathsf{t}, \mathsf{t}^\prime$ and $\mathsf{r}, \mathsf{r}^\prime$ denoting the $2 \times 2$ transmission and reflection blocks. The transmission coefficient is now obtained as $T(E)={\rm}{tr}[\mathsf{t}_E^\dagger \mathsf{t}^{}_E]={\rm}{tr}[ \mathsf{t}^{\prime\,  \dagger}_E  \mathsf{t}^{\prime}_E]$.\\\\

{\it Generalization to multiple interfaces.} 
It is straightforward to generalise the same approach to the case of $N$ interfaces. Let us consider an interface located at $x_j$ separating a region  with RSOC $\alpha_j$ (on the left side) from a region with RSOC $\alpha_{j+1}$ on the right, as depicted in Fig.\ref{Fig10}. Similarly to the   single interface problem [see Eqs.(\ref{Ansatz-single-1}) and (\ref{Ansatz-single-2})], on each side of the interface the electron field operator $\hat{\Psi}_E(x)$ is written as an expansion of (at most) four modes, which can be propagating or evanescent, depending on the specific regime of that region and on the energy range, according to Tables \ref{table-Rashba}, \ref{table-Weak-Zeeman}, \ref{table-Strong-Zeeman}.  Denoting by $\hat{c}_{E}^{(j)}$, $\hat{d}_{E}^{(j)}$, $\hat{f}_{E}^{(j)}$, and $\hat{g}_{E}^{(j)}$ the operators related to the above  four modes on the $j$-th region and the boundary conditions (\ref{bc-single-interface}) can be rewritten in a matrix form
\begin{equation}
    M_E^{(j)}(x_j) \Vector{\hat{c}_{E}^{(j)}\vspace{3pt} \\ \hat{d}_{E}^{(j)}\vspace{3pt} \\ \hat{f}_{E}^{(j)}\vspace{3pt} \\ \hat{g}_{E}^{(j)}} = M_E^{(j+1)}(x_j) \Vector{\hat{c}_{E}^{(j+1)}\vspace{3pt} \\ \hat{d}_{E}^{(j+1)}\vspace{3pt} \\ \hat{f}_{E}^{(j+1)}\vspace{3pt} \\ \hat{g}_{E}^{(j+1)}}\label{Wtot}
\end{equation}
\begin{figure} 
        \includegraphics[width=12cm]{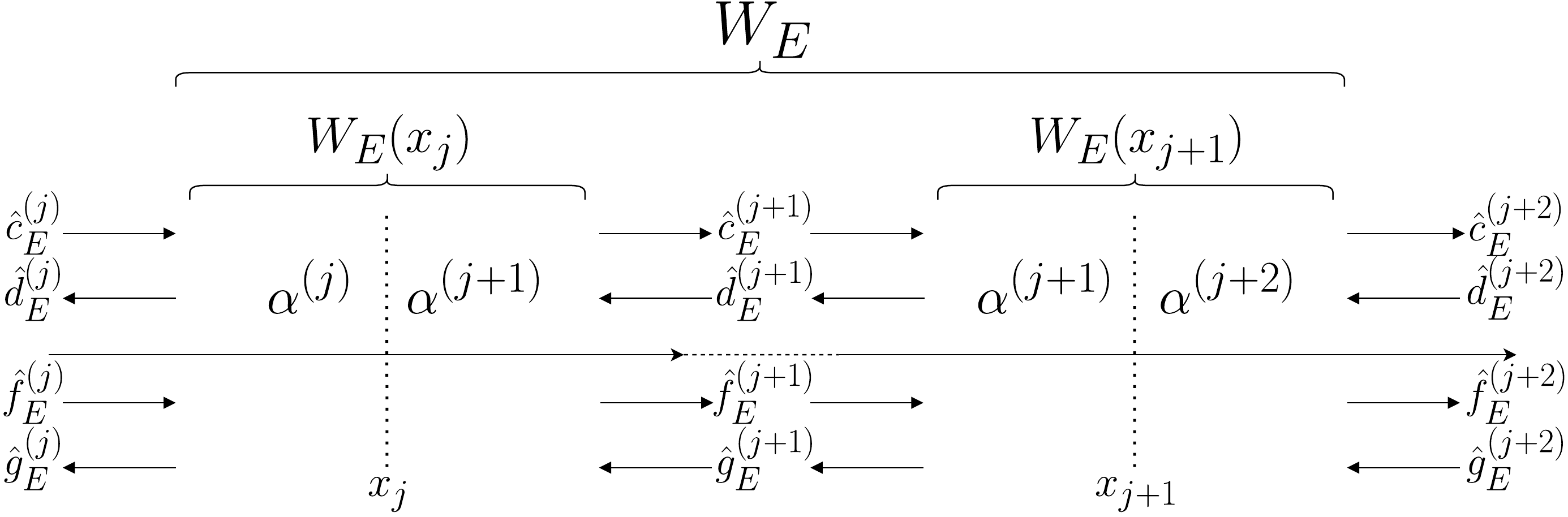}
    \caption{ \small  Generalization of the transfer matrix to multiple interfaces. The total transfer matrix is computed as the product of the transfer matrices across each interface.}
    \label{Fig10}
\end{figure}
The transfer matrix across the interface at $x_j$ is obtained as $W_E(x_j) =  [ M_E^{(j+1)}(x_j)]^{-1} M_E^{(j)}(x_j)$. Then, by observing that the operators on right of   the interface $x_j$ are the very same  operators as the ones on the left of the interface at $x_{j+1}$ (see Fig.\ref{Fig10}),     the total transfer matrix $W_E$ across the entire inhomogeneous scattering region 
\begin{equation}
    \Vector{\hat{c}_{E}^{(N)}\vspace{3pt} \\ \hat{d}_{E}^{(N)}\vspace{3pt} \\ \hat{f}_{E}^{(N)}\vspace{3pt} \\ \hat{g}_{E}^{(N+1)}} = W_E  \Vector{\hat{c}_{E}^{(0)}\vspace{3pt} \\ \hat{d}_{E}^{(0)}\vspace{3pt} \\ \hat{f}_{E}^{(0)}\vspace{3pt} \\ \hat{g}_{E}^{(0)}} \label{Transfer-single-interface}
\end{equation}
can be easily found as  the product $W_E = \prod_{j=0}^{N-1} W_E(x_{N - j-1})$. For instance, for the  doubly gated NW problem of Sec.\ref{sec-5}, the total transfer matrix is the product $W_E=W_E(x_1)W_E(x_0)$ of the two single-interface transfer matrices and contains the distance $d=x_1-x_0$ between the two interfaces at $x_0$ and $x_1$. 
Finally, one observes that in the outer region  $j=0$ (left lead) the modes propagating rightwards (leftwards) are incoming to (outgoing from) the scattering region   and can thus be relabelled as $\hat{a}^{(L)}_{\beta E}$ ($\hat{b}^{(L)}_{\beta E}$). Moreover,  in order to ensure the normalizability of the state, only the evanescent mode decaying to the left is admitted, like in Eq.(\ref{Ansatz-single-1}). Effectively, this means that in Eq.(\ref{Wtot}),  the operator related to the mode diverging on the left has to be set to zero, like in Eq.(\ref{Transfer-single-interface-1}). Similarly, in the outer region $j=N$ (right lead) the modes propagating leftwards (rightwards) are incoming (outgoing) and are relabelled as $\hat{a}^{(R)}_{\beta E}$ ($\hat{b}^{(R)}_{\beta E}$), while only the evanescent mode  decaying on the right, if any,  is admitted. Implementing these aspects in Eq.(\ref{Wtot}), and re-expressing the outgoing mode operators $\hat{b}^{(L)}_{\beta E},\hat{b}^{(R)}_{\beta E}$ in terms of the incoming ones $\hat{a}^{(L)}_{\beta E},\hat{a}^{(R)}_{\beta E}$, one finds the Scattering Matrix similarly to what was done in Eqs.(\ref{S-matrix-single-1}) and Eqs.(\ref{S-matrix-single-2}) for the single interface case.
 
\end{widetext}

 

\begin{thebibliography}{994}
\bibitem{dassarma_2010} R. M. Lutchyn, J. D. Sau, and S. Das Sarma, Phys. Rev. Lett. {\bf 105}, 077001 (2010).
\bibitem{vonoppen_2010} Y. Oreg,  G. Refael,  and F. von Oppen Phys. Rev. Lett. {\bf 105}, 177002 (2010). 
\bibitem{alicea_review} J.Alicea, Rep. Prog. Phys. {\bf 75}  076501  (2012).
\bibitem{aguado_review} R. Aguado, La Rivista del Nuovo Cimento, {\bf 40}, 523 (2017).
\bibitem{prada-review}  E. Prada, P. San-Jos\'e, M.W.A. de Moor, A. Geresdi, E.J.H. Lee, J. Klinovaja, D. Loss, J. Nyg{\aa}rd, R. Aguado, and L.P. Kouwenhoven, Nat. Rev. Phys. {\bf 2}, 575 (2020). 
\bibitem{valkov-review} V.V. Val'kov, M.S. Shustin, S.V. Aksenov, A.O. Zlotnikov, A.D. Fedoseev, V.A. Mitskan, M.Y. Kagan, Phys. Usp. {\bf 65}, 2 (2022).

\bibitem{kouwenhoven_2012} V. Mourik,  K. Zuo,  S.M. Frolov,  S.R. Plissard,  E.P.A.M. Bakkers, L.P. Kouwenhoven, Science  {\bf 336}, 1003 (2012).
\bibitem{liu_2012} L.P. Rokhinson, X. Liu, and J. K. Furdyna,   Nat. Phys.  {\bf 8}, 795 (2012).
\bibitem{heiblum_2012} A. Das, Y. Ronen, Y. Most, Y. Oreg, M. Heiblum, and H. Shtrikman,  Nat. Phys.  {\bf 8}, 887 (2012).
\bibitem{xu_2012} M.T. Deng,  C. L. Yu,  G. Y. Huang,  M. Larsson, P. Caroff,  and H. Q. Xu,  Nano Lett.  {\bf 12}, 6414 (2012).
\bibitem{defranceschi_2014} E.J.H. Lee, X. Jiang, M. Houzet, R. Aguado, C. M. Lieber, and S. De Franceschi,   Nat. Nanotech. {\bf 9}, 79 (2014).
\bibitem{marcus_2016} S.M. Albrecht, A. P. Higginbotham, M. Madsen, F. Kuemmeth, T. S. Jespersen, J. Nyg{\aa}rd, P. Krogstrup, and C. M. Marcus,   Nature  {\bf 531}, 206 (2016).
\bibitem{marcus_science_2016} M.T. Deng, S. Vaitiekenas,  E. B. Hansen,  J. Danon,  M. Leijnse,  K. Flensberg,  J. Nyg{\aa}rd,  P. Krogstrup, and  C. M. Marcus,  Science  {\bf 354}, 1557 (2016).
\bibitem{kouwenhoven_2018} \"O. G\"ul , H. Zhang, J.D.S. Bommer, M.W. A. de Moor, D. Car, S.R. Plissard, E.P.A.M. Bakkers, A. Geresdi, K. Watanabe, T. Taniguchi, and L.P. Kouwenhoven, Nat. Nanotech. {\bf 13}, 192 (2018).

\bibitem{kallaher_2010} R.L. Kallaher, J.J. Heremans, N. Goel, S.J. Chung, and M.B. Santos, Phys. Rev. B {\bf 81}, 035335 (2010)
\bibitem{kouwenhoven_nanolett_2013} I. van Weperen, S.R. Plissard, E.P.A.M. Bakkers, S.M. Frolov, and Leo P. Kouwenhoven, Nano Lett. {\bf 13}, 387 (2013). 
\bibitem{kouwenhoven_nanolett_2016} J. Kammhuber, M.C. Cassidy, H. Zhang, \"O G\"ul, F. Pei, M.W.A. de Moor, B. Nijholt, K. Watanabe, T. Taniguchi, D. Car, S.R. Plissard, E.P.A.M. Bakkers, and L.P. Kouwenhoven, Nano Lett. {\bf 16}, 3482 (2016).  
\bibitem{shaepers_nanolett_2016} S. Heedt, W. Prost, J. Schubert, D. Gr\"utzmacher, and Th. Sch\"apers, Nanolett {\bf 16}, 3116 (2016).
\bibitem{xu_2016} S. Li, N. Kang, D.X. Fan, L.B. Wang, Y.Q. Huang, P. Caroff, H.Q. Xu,  Sci.Rep {\bf 6}, 24822 (2016).  
\bibitem{kouwenhoven_2017a} H. Zhang, \"O G\"ul, S. Conesa-Boj, M. P. Nowak, M. Wimmer, K. Zuo, V. Mourik, F. K. de Vries, J. van Veen, M.W.A. de Moor, J.S. Bommer, D.J. van Woerkom, D.Car, S.R. Plissard, E.P.A.M. Bakkers, M.Quintero-P\'erez, M.C. Cassidy, S. Koelling, S. Goswami, K. Watanabe, T. Taniguchi, and L. P. Kouwenhoven, Nat. Comm. {\bf 8}, 16025 (2017).
\bibitem{kouwenhoven_2017b} E.M.T. Fadaly, H. Zhang, S. Conesa-Boj, D. Car, \"O. G\"ul, S.R. Plissard, R. L.M. OphetVeld, S. K\"olling, L.P. Kouwenhoven, and E. P. A. M. Bakkers, Nanolett. {\bf 17}, 6511 (2017). 
\bibitem{DeFranceschi_2018} J. C. Estrada Saldana, Y.-M. Niquet, J.-P. Cleuziou, E.J. H. Lee, D. Car, S.R .Plissard, E. P.A.M.Bakkers, and S. De Franceschi, Nanolett {\bf 18}, 2282 (2018). 
\bibitem{kouwenhoven_2019a} P. Aseev, G.Wang, L. Binci, A. Singh, S. Marti-Sanchez, M. Botifoll, L.J. Stek, A. Bordin, J.D. Watson, F. Boekhout,  D. Abel, J.Gamble, K. Van Hoogdalem, J. Arbiol, L.P. Kouwenhoven, G. de Lange, and Ph. Caroff,  Nanolett {\bf 19}, 9102 (2019).
\bibitem{kouwenhoven_2019b} G. Badawy, S. Gazibegovic, F. Borsoi, S. Heedt, C.A. Wang, S.Koelling, M.A. Verheijen, L. P. Kouwenhoven, E.P.A.M. Bakkers, Nanolett {\bf 19}, 3575 (2019). 


\bibitem{schaepers_2020}  P. Zellekens, N. Demarina, J. Jan{\ss}en, T, Rieger, M. I. Lepsa, P. Perla, G.Panaitov, H. L\"uth, D. Gr\"utzmacher,  and Th. Sch\"apers, Semicond. Sci. Technol. {\bf 35}, 085003 (2020). 

\bibitem{giazotto_2019} A. Iorio, M. Rocci, L. Bours, M. Carrega, V. Zannier, L. Sorba, S. Roddaro, F. Giazotto, and E. Strambini, Nanolett. {\bf 19}, 652 (2019).

\bibitem{bakkers_2012} S. R.Plissard, D.R. Slapak, M.A.Verheijen, M. Hocevar, G.W.G. Immink, I. van Weperen, S. Nadj-Perge,  S.M. Frolov, L.P. Kouwenhoven, and E.P.A. M.Bakkers, Nanolett {\bf 12}, 1794 (2012).
\bibitem{philipose_2019} A.P. Singh, K. Roccapriore, Z. Algarni, R. Salloom, T.D. Golden, and U. Philipose, Nanomat. {\bf 9}, 1260 (2019).


\bibitem{krogstrup_2020} Y. Liu, S. Vaitiekenas, S. Mart\'i-S\'anchez, Ch. Koch, S. Hart, Z. Cui, Th. Kanne, S.A. Khan, R.Tanta, S. Upadhyay, M. Espi{\~{n}}eira Cachaza, C.M. Marcus, J. Arbiol, K.A. Moler, and P. Krogstrup, Nanolett. {\bf 20}, 456 (2020). 
\bibitem{sasaki_2013} S. Sasaki,  K. Tateno, G. Zhang, H. Suominen, Y. Harada, S. Saito, A. Fujiwara, T. Sogawa, and K. Muraki,  Appl. Phys. Lett. {\bf 103}, 213502 (2013).
\bibitem{micolich} A. M. Burke, D. J. Carrad, J. G. Gluschke, K. Storm, S. Fahlvik Svensson, H. Linke, L. Samuelson, and A. P. Micolich, Nano Lett. {\bf 15}, 2836 (2015).
\bibitem{sasaki_2017} K. Takase, Y. Ashikawa, G. Zhang, K. Tateno, and S. Sasaki, Sci. Rep. {\bf 7}, 930 (2017).
\bibitem{das_2019} S.R. Das,   in   {\it Nanoelectronics: Devices, Circuits and Systems}, edited by B.K. Kaushik  (Elsevier, Amsterdam, 2019), Chap.11, p. 355.
\bibitem{guo_2021} Y. Yin, Z.  Zhang, H.  Zhong, C. Shao, X. Wan, C. Zhang, J.  Robertson, and Y. Guo,  ACS Appl. Mat.  and Interfaces  {\bf 13}, 3387 (2021).
\bibitem{sasaki_2021} K. Takase, K.  Tateno, and S. Sasaki,   Appl. Phys. Lett. {\bf 119}, 013102 (2021). 
\bibitem{gao_2012} D. Liang and X. P.A. Gao, Nano Lett. {\bf 12}, 3263 (2012).
\bibitem{slomski_NJP_2013} B. Slomski, G. Landolt, S. Muff, F. Meier, J. Osterwalder, and J.H. Dil, New J. Phys. {\bf 15}, 125031  (2013).
\bibitem{wimmer_2015} I. van Weperen, B. Tarasinski, D. Eeltink, V. S. Pribiag, S. R. Plissard, E.P.A.M. Bakkers,  L.P. Kouwenhoven,  and M. Wimmer, Phys. Rev. B {\bf 91},   201413(R) (2015).
\bibitem{bercioux_review} D. Bercioux, and P. Lucignano, Rep. Progr. Phys. {\bf 78},  106001 (2015).
\bibitem{nygaard_2016} Z. Scher\"ubl,  G. F\"ul\"op, M. H. Madsen, J. Nyg{\aa}rd, and S. Csonka, Phys. Rev. B {\bf 94}, 035444 (2016).
\bibitem{sherman_2016} J. R. Bindel, M. Pezzotta, J. Ulrich, M. Liebmann, E. Y. Sherman, and M. Morgenstern, Nat. Phys. {\bf 12}, 920 (2016).
\bibitem{tokatly_PRB_2017} J. Borge and I. V. Tokatly, Phys. Rev. B {\bf 96}, 115445 (2017).
\bibitem{loss_2018} Ch. Kloeffel,  M. J. Ran\v{c}i\'{c}, and D. Loss, Phys. Rev. B {\bf 97}, 235422 (2018).
\bibitem{goldoni_2018} P. W\'ojcik, A. Bertoni, and G. Goldoni, Phys. Rev. B {\bf  97}, 165401 (2018).
\bibitem{gao-review} K. Premasiri  and X. P. A. Gao, J. Phys. Cond. Matt. {\bf 31},  193001 (2019).
\bibitem{lau_2021} D. Shcherbakov, P. Stepanov, S. Memaran, Y. Wang, Y. Xin, 
J. Yang, K. Wei, R. Baumbach, W. Zheng, K. Watanabe, T. Taniguchi, M. Bockrath, D. Smirnov, T. Siegrist2, W. Windl, 
L. Balicas, C.N. Lau, Sci. Adv. {\bf 7}, eabe2892 (2021).

\bibitem{nitta-frustaglia} F. Nagasawa, A.A. Reynoso, J.P. Baltan\'as, D. Frustaglia, H. Saarikoski, and J. Nitta, Phys. Rev. B {\bf 98}, 245301 (2018).
\bibitem{kaindl_2005} O. Krupin, G. Bihlmayer, K. Starke, S. Gorovikov,  J. E. Prieto,  K. D\"obrich, S. Bl\"ugel, and G. Kaindl, Phys. Rev. B {\bf 71}, 201403(R) (2005).
\bibitem{slomski_2013} B. Slomski, G. Landolt, S. Muff, F. Meier, J. Osterwalder, and J.H. Dil, New J. Phys. {\bf 15}, 125031 (2013).
\bibitem{tsai_2018} H. Tsai, S. Karube, K. Kondou, N. Yamaguchi, and Y. Otani, Sci. Rep. {\bf 8}, 5564 (2018).
\bibitem{wang-fu_2016} W. Wang, X.M. Li, J.Y. Fu, J. Magn. Magn. Mat. {\bf 411}, 84 (2016).
\bibitem{johannesson-japaridze_2011} M. Malard, I. Grusha, G.I. Japaridze, and H. Johannesson, Phys. Rev. B {\bf 84}, 075466 (2011). 
\bibitem{streda} P. Str\v{e}da, and P. \v{S}eba, Phys. Rev. Lett. {\bf 90}, 256601 (2003).
\bibitem{depicciotto_2010} C.H.L. Quay, T.L. Hughes, J.A. Sulpizio, L.N. Pfeiffer, K.W. Baldwin, K. W. West, D. Goldhaber-Gordon, and R. de Picciotto, Nat. Phys {\bf 6}, 336 (2010).
\bibitem{loss_PRB_2011} C. Kloeffel,  M. Trif, and D. Loss,  Phys. Rev. B {\bf  84}, 195314 (2011). 
\bibitem{lutchyn_2012} M. Cheng, and R. M. Lutchyn, Phys. Rev. B {\bf 86}, 134522 (2012).
\bibitem{loss_PRB_2017}  P. Szumniak, D. Chevallier, D. Loss,  and J. Klinovaja, Phys. Rev. B {\bf 96}, 041401(R) (2017).

\bibitem{loss_EPJB_2015} J. Klinovaja and D. Loss, Eur. Phys. J. B {\bf 88}, 62 (2015).
\bibitem{rossi-dolcini-rossi_EPJ} L. Rossi, F. Dolcini, and F. Rossi, Eur. Phys. J. Plus {\bf 135} 597 (2020).


\bibitem{rossi-dolcini-rossi_2020} L. Rossi, F. Dolcini, and F. Rossi, Phys. Rev. B {\bf 101}, 195421 (2020).


\bibitem{sherman_PRL_2007} M.M. Glazov, and E. Ya. Sherman, Phys. Rev. Lett. {\bf 107}, 156602 (2007).
\bibitem{brataas_2007} Y. Tserkovnyak, B.I. Halperin, A.A. Kovalev, A. Brataas, Phys. Rev. B {\bf 76}, 085319 (2007).
\bibitem{sherman_PRB_2013} A.F. Sadreev  and E. Ya. Sherman, Phys. Rev. B {\bf 88}, 115302 (2013).
\bibitem{sherman_PRB_2018} S. Kud{\l}a, A. Dyrda{\l}, V.K. Dugaev, E. Ya. Sherman, and J. Barna\'{s}, Phys. Rev. B {\bf 97}, 245307 (2018).
\bibitem{rashba_2004} E.I. Rashba, Phys. Rev. B {\bf 70}, 161201 (2004).


\bibitem{dolcini-rossi_2018} F. Dolcini and F. Rossi, Phys. Rev. B {\bf 98}, 045436 (2018).
\bibitem{sanchez_2006} D. S\'anchez, L. Serra, Phys. Rev. B {\bf 74} 153313 (2006).



\bibitem{note-bc} Note that the boundary conditions (\ref{bc-single-interface}), when rewritten in terms of $\hat{\Psi}^\prime$ through Eq.(\ref{gauge-transf}), acquire the standard  form of continuity of the spinor field $\hat{\Psi}^\prime$ and its derivative $\partial_x\hat{\Psi}^\prime$. 
\bibitem{datta_book} S. Datta {\it Electronic Transport in Mesoscopic Systems} (Cambridge University Press, Cambridge, 1995).
\bibitem{sanchez_2008} D. S\'anchez, L. Serra,  and M.-S. Choi, Phys. Rev. B {\bf 77},    035315 (2008).
\bibitem{aguado_2015} J. Cayao, E. Prada,  P. San-Jose,  and R. Aguado, Phys. Rev. B {\bf 91}, 024514 (2015).
\bibitem{rainis-loss_PRB_2014} D. Rainis and D. Loss, Phys. Rev. B {\bf 90}, 235415 (2014).


\bibitem{nota-config} Although the two  configurations (Zeeman, Rashba) and (Rashba, Zeeman) with exchanged values, are a priori physically different because of the    magnetic field applied along the NW axis, they actually yield  the same transmission coefficient $T(E)$, for the latter turns out to be independent of the sign of the  magnetic field along the NW.  This  can be seen by performing the gauge transformation $\hat{\Psi}\rightarrow \hat{\Psi}^{\prime}=\sigma_z \hat{\Psi}$ on the electron field in the Hamiltonian Eq.(1), which maps the case of Zeeman field $+h_\perp$ into the case $-h_\perp$. 




 \bibitem{nota-su-bs} Bound states localized at the interface may exist, but they are energetically  located below the continuum, not in the magnetic gap range (see Ref.[\onlinecite{rossi-dolcini-rossi_2020}]). 

\bibitem{gogin_2022} L. Gogin, L. Rossi, F. Rossi, and F. Dolcini, New J. Phys. {\bf 24}, 053045 (2022).
\bibitem{sasaki-NTT-review} K. Takase, G. Zhang, K. Tateno, and S. Sasaki, NTT Technical Review, {\bf 17}, 56 (2019). 


\bibitem{peres} N.M.R. Peres, J. Phys.: Condens. Matter {\bf 21}, 095501 (2009).
 

\bibitem{malshukov_2003} A. G. Mal'shukov, C. S. Tang, C. S. Chu, and K. A. Chao, Phys. Rev. B {\bf 68}, 233307 (2003).
\bibitem{malshukov_2005} C. S. Tang, A. G. Mal'shukov,  and K. A. Chao, Phys. Rev. B {\bf 71}, 195314 (2005).
\bibitem{dolcini_2012} F. Dolcini, Phys. Rev. B {\bf 85}, 033306 (2012). 
\bibitem{loss_2016} J.  Klinovaja, P. Stano, and  D. Loss, Phys. Rev. Lett. {\bf 116}, 176401 (2016).
 
\end{thebibliography}
\end{document}